# Scaling of the thermally induced sign inversion of longitudinal spin Seebeck effect in a compensated ferrimagnet: Role of magnetic anisotropy

*Amit Chanda, Noah Schulz, Manh-Huong Phan\* and Hari Srikanth\**

Department of Physics, University of South Florida, Tampa, Florida 33620, USA

E-mail: phanm@usf.edu; sharihar@usf.edu

*Christian Holzmann, Johannes Seyd and Manfred Albrecht\**

Institute of Physics, University of Augsburg, 86159 Augsburg, Germany

E-mail: manfred.albrecht@physik.uni-augsburg.de

We report on a systematic investigation of the longitudinal spin Seebeck effect (LSSE) in a GGG($Gd_3Ga_5O_{12}$)/GdIG($Gd_3Fe_5O_{12}$)/Pt film series exhibiting an in-plane magnetic easy axis with a compensation temperature ($T_{Comp}$) that decreases from 270 to 220 K when decreasing GdIG film thickness from 272 to 31 nm, respectively. For all the films, the LSSE signal flips its sign below $T_{Comp}$. We demonstrate a universal scaling behavior of the temperature dependence of LSSE signal for our GdIG films around their respective $T_{Comp}$. Additionally, we demonstrate LSSE in a 31 nm GdIG film grown on a lattice-mismatched GSGG ($Gd_3Sc_2Ga_3O_{12}$) substrate that exhibits an out-of-plane magnetic easy axis at room temperature. However, this sample reveals a spin reorientation transition where the magnetic easy axis changes its orientation to in-plane at low temperatures. We observed a clear distinction in the LSSE signal for the GSGG/GdIG(31 nm)/Pt heterostructure, relative to GGG/GdIG(31nm)/Pt showing an in-plane magnetic easy axis. Our findings underscore a strong correlation between the LSSE signal and the orientation of magnetic easy axis in compensated ferrimagnets and opens the possibility to tune LSSE through effective anisotropy.

# 1. Introduction

To address the global need for sustainable energy resources, thermoelectric effects, especially the longitudinal Seebeck effect, which is also recognized as the charge Seebeck effect (CSE), has been known for the past few decades as a clean and efficient way to harvest electrical energy from renewable thermal energy sources. The efficiency of a conventional CSE based thermoelectric generator is described by the thermoelectric figure of merit (FoM): $Z_{CSE}T = \frac{S^2}{\kappa\rho}T$, where $S$, $\kappa$, and $\rho$ are the Seebeck coefficient, the thermal conductivity, and the electrical resistivity of the material, respectively.[1] Enhancement of the thermoelectric conversion efficiency requires a material with a higher $S$-value but lower values of $\kappa$ and $\rho$, which is restricted by the Wiedemann-Franz law for isotropic conductors. In analogy to the CSE, the spin Seebeck effect (SSE) was discovered in 2008,[2] which represents a thermally driven spin to charge current conversion process and is typically observed in a ferro(i)magnet (FM)/nonmagnetic metal (NM) bilayer structure. The FoM for such an FM/NM hybrid structure is defined as $Z_{SSE}T = \frac{(\theta_{SH}Q_{SSE})^2}{\kappa_{FM}\rho_{NM}}T$, where $\theta_{SH}$, $Q_{SSE}$, $\kappa_{FM}$, and $\rho_{NM}$ are the spin-Hall angle of the NM, the SSE coefficient of the FM, the thermal conductivity of the FM, and the electrical resistivity of the NM, respectively.[3] Thus, the efficiency of a SSE-based thermoelectric generator is not limited by the Wiedemann-Franz law and can be optimized by controlling multiple degrees of freedom. Depending on the experimental geometry, SSE can be categorized into the longitudinal and transverse SSE configurations.[4] Longitudinal SSE (LSSE) leads to the generation of incoherent terahertz magnon excitations in a FM insulator parallel to the direction of an applied temperature gradient.[5] Unlike charge currents, the magnon spin current in a magnetic insulator with frozen electronic degrees of freedom eliminates the possibility of energy dissipation due to ohmic losses.[6] This thermally generated magnon spin current is electrically realized by placing a thin layer of heavy metal with strong spin-orbit coupling (*e.g.*, Pt, Ta) in close contact with the magnetic insulator which converts the pumped spin current from the magnetic insulator into charge current via the inverse spin Hall effect (ISHE).[7]

Past few years have witnessed that insulating rare-earth iron garnets (REIGs) and ferrites are the most suitable class of materials from the perspective of pure spin current source as they do not cause contamination of the thermally generated spin current by the electronic degrees of freedom.[8][9][10][11][12] Although the ferrimagnetic insulator $Y_3Fe_5O_{12}$ (YIG) is known as the ideal system for LSSE mainly because of its ultra-low Gilbert damping,[13][14][15][16][17][18][19][20][21] other members of REIG family did not receive much attention till date. Geprägs *et al*.[22] observed an anomalous temperature dependence of the LSSE signal in $Gd_3Fe_5O_{12}$ (GdIG) which was completely different from that observed for YIG. GdIG is a compensated ferrimagnetic insulator and consists of three magnetic sublattices: two antiparallelly aligned $Fe^{3+}$ sublattices and one $Gd^{3+}$ sublattice.[22] The antiferromagnetic exchange coupling between the $Gd^{3+}$ and $Fe^{3+}$ ions is very weak, which causes strong temperature sensitivity of the $Gd^{3+}$ sublattice magnetization compared to the $Fe^{3+}$ sublattices.[23][24] Moreover, the $Gd^{3+}$ sublattice magnetization increases drastically at low temperatures and overcomes the magnetization of the $Fe^{3+}$ sublattices. This leads to the existence of a magnetic compensation temperature ($T_{Comp}$) close to room temperature at which the net magnetization becomes zero.[22] The LSSE signal was shown to flip its sign at $T_{Comp}$ because of reorientation of sublattice magnetizations. Interestingly, the LSSE signal shows a second sign change at low temperatures ($\approx$ 80 K), which is understood in terms of enhanced spin injection efficiency of the $Gd^{3+}$ moment dominated gapless magnon mode ($\alpha$ mode) compared to the $Fe^{3+}$ moment dominated gapped magnon mode ($\beta$ mode).[22][24] Later, Yagmur *et al*.[25] investigated the temperature dependences of LSSE and spin Peltier effect (SPE) in a GdIG/Pt junction using the lock-in thermography technique and observed the similar changes in sign of both LSSE and SPE around $T_{Comp}$ of GdIG. Such fascinating observation of multiple sign reversals of the LSSE signal in GdIG/Pt bilayer motived us to address an emerging fundamental question in this work: *Is there a general trend for the temperature dependence of LSSE in compensated ferrimagnets*? Such magnon mode-selective thermo-spin transport has been observed in a few members of the REIG family,[22][24][26] but it is rare amongst other compensated ferrimagnets. To address our quest for a universal trend of LSSE in compensated ferrimagnets around their magnetic compensation temperatures, we focused on the first sign change around $T_{Comp}$ and investigated LSSE within a temperature window close to $T_{Comp}$. Another extraordinary characteristic of the insulating iron garnet thin films is that the orientation of magnetic easy axis can simply be changed from in-plane (IP) to out-of-plane (OOP) just by tuning the film strain induced by the substrate.[27] Previously, our group demonstrated a strong correlation between LSSE signal and magnetic anisotropy in YIG.[19] However, there is no previous report on the influence of the orientation of magnetic easy axis on the LSSE signal, especially in compensated ferrimagnets.

In this paper, we have performed a systematic investigation of LSSE in GGG/GdIG(*t*)/Pt(5nm) heterostructures with 5 different thicknesses ranging between 31 and 272 nm. All GdIG films possess an IP magnetic easy axis, and $T_{Comp}$ decreases with decreasing film thickness. The LSSE signal for all the heterostructures flips its sign below $T_{Comp}$. Using a proposed rescaling method, we have demonstrated a "universal scaling" behavior of the temperature dependence of LSSE signal around the respective $T_{Comp}$. Additionally, we have investigated LSSE in a 31 nm GdIG film grown on a lattice-mismatched GSGG substrate that exhibits an OOP magnetic easy axis at room temperature and observed a clear distinction in the magnetic field dependent LSSE signal, relative to GGG/GdIG(31nm)/Pt heterostructure exhibiting shape anisotropy-driven IP magnetic easy axis. This highlights a strong correlation between the LSSE signal and the magnetic anisotropy in compensated ferrimagnets.

## 2. Results and Discussion

### 2.1. Structural Characterization

**Figure 1**(a) shows the X-ray diffraction (XRD) $\theta - 2\theta$ diffractograms of the (444) planes for gadolinium iron garnet (GdIG) films with different film thicknesses grown on $Gd_3Ga_5O_{12}$ (GGG) and $Gd_3Sc_2Ga_3O_{12}$ (GSGG) substrates (**Figure 1**(b)). The sharp Bragg peaks associated with the substrate are visible at 51.08° and 50.28° for GGG and GSGG, respectively. The substrate choice and the film thickness show a strong influence on the observed film structure. The Bragg (444) peaks associated with the GdIG films are visible at lower angles than the bulk reflection for films grown on GGG, indicating tensile out-of-plane strain, but at a higher angle for the films on GSGG, suggesting out-of-plane compressive strain, which is required to achieve perpendicular magnetic anisotropy (PMA). The bulk (444) reflection is marked by a dashed line.

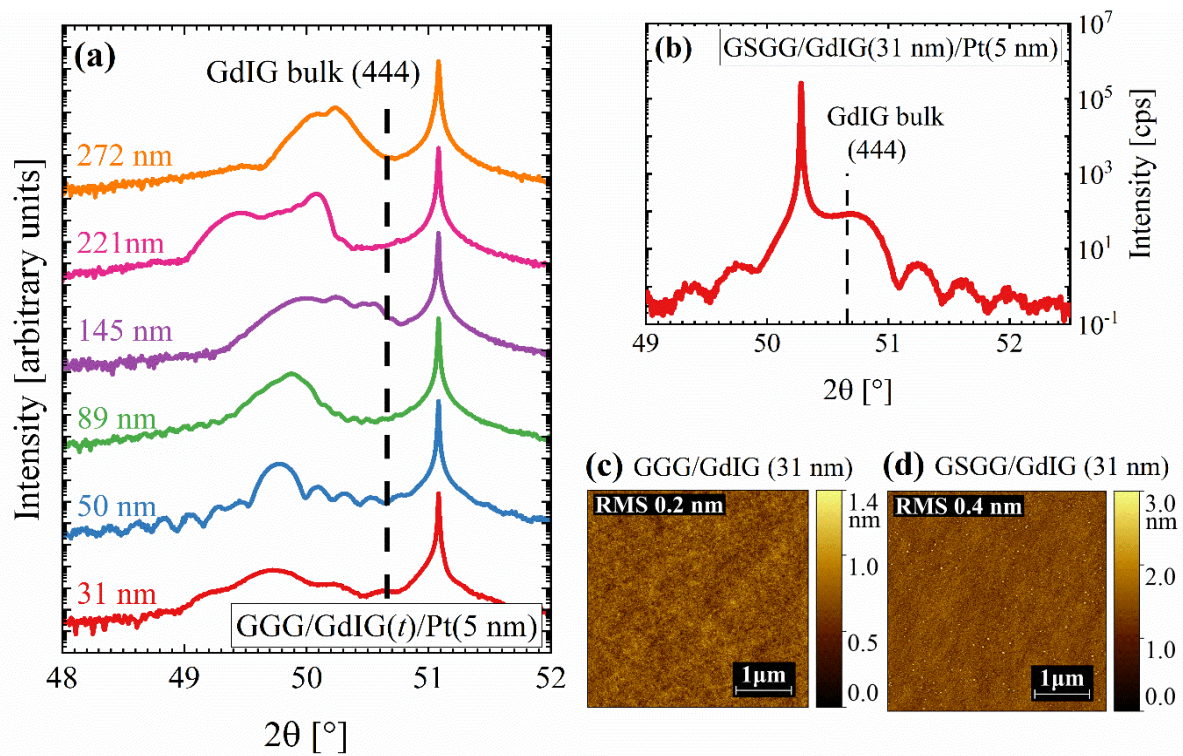

**Figure 1.** (a) $\theta - 2\theta$ X-ray diffractograms (XRD) of the GGG/GdIG ($t$)/Pt films with different film thicknesses $t$ and (b) GSGG/GdIG(31 nm)/Pt, showing the substrate and film (444) reflections. (c) Film morphology as visible in atomic force microscopy (AFM) for a film grown on GGG and (d) on GSGG.

While the thinnest film shows the highest amount of strain, a gradual relaxation towards the bulk structure with increasing film thickness is observed, visible by the increasing Bragg angle in **Figure 1**(a). The thickest film of 272 nm thickness seems to be not yet fully relaxed, as it shows a Bragg peak at a smaller angle than the bulk reflection. However, a slight off-stoichiometry present in the thin film samples could create an offset between the relaxed film (444) Bragg peak and the bulk (444) reflection. Additionally, several individual peaks close to each other are visible for the thicker films. This might be caused by slightly varying off-stoichiometry introduced into the film during the long growth process, however no apparent impact on the magnetic properties of the thin film is observed. Recently, the peak splitting has been related to differently strained layers in samarium iron garnet (SmIG) thin films due to the relaxation of the film unit cell with increasing film thickness[30]. Additionally, for the thinner films, Laue oscillations are apparent. While these oscillations indicate single-crystalline order with smooth interfaces, the Laue peaks move closer with increasing film thickness and are therefore hardly visible for the thicker films. Simulating the XRD signal by the addition of a substrate contribution in the shape of a pseudo-Voigt function and a film contribution given by the Laue oscillations yields the Bragg peak position and film thickness.[28] However, as this evaluation is not possible for the thicker film, additional spectroscopic ellipsometry measurements were conducted to analyze the film thickness. Note that all the films show a smooth surface morphology before depositing the Pt strip, crucial for generating an inverse spin-Hall current at the GdIG/Pt interface. A low root-mean-square roughness below 0.5 nm is achieved for all films, as visible in the atomic force microscopy (AFM) images shown for the 31 nm thick GdIG films on GGG in **Figure 1**(c) and GSGG in **Figure 1**(d). AFM images for the remaining films are shown in the **supplementary information (Figure S1 and S2)**.

## 2.2. Magnetization and Effective Magnetic Anisotropy

Main panels of **Figure 2**(a)-(f) display the temperature dependence of saturation magnetization, $M_S$, obtained from $M(H)$ hysteresis loops measured in in-plane configuration on the G(S)GG/GdIG($t$) films with $t$ = 272-31 nm. Temperature dependence of $M_S$ and $M(H)$ for the GGG/GdIG(145 nm) film are shown in the **supplementary information (Figure S4).**

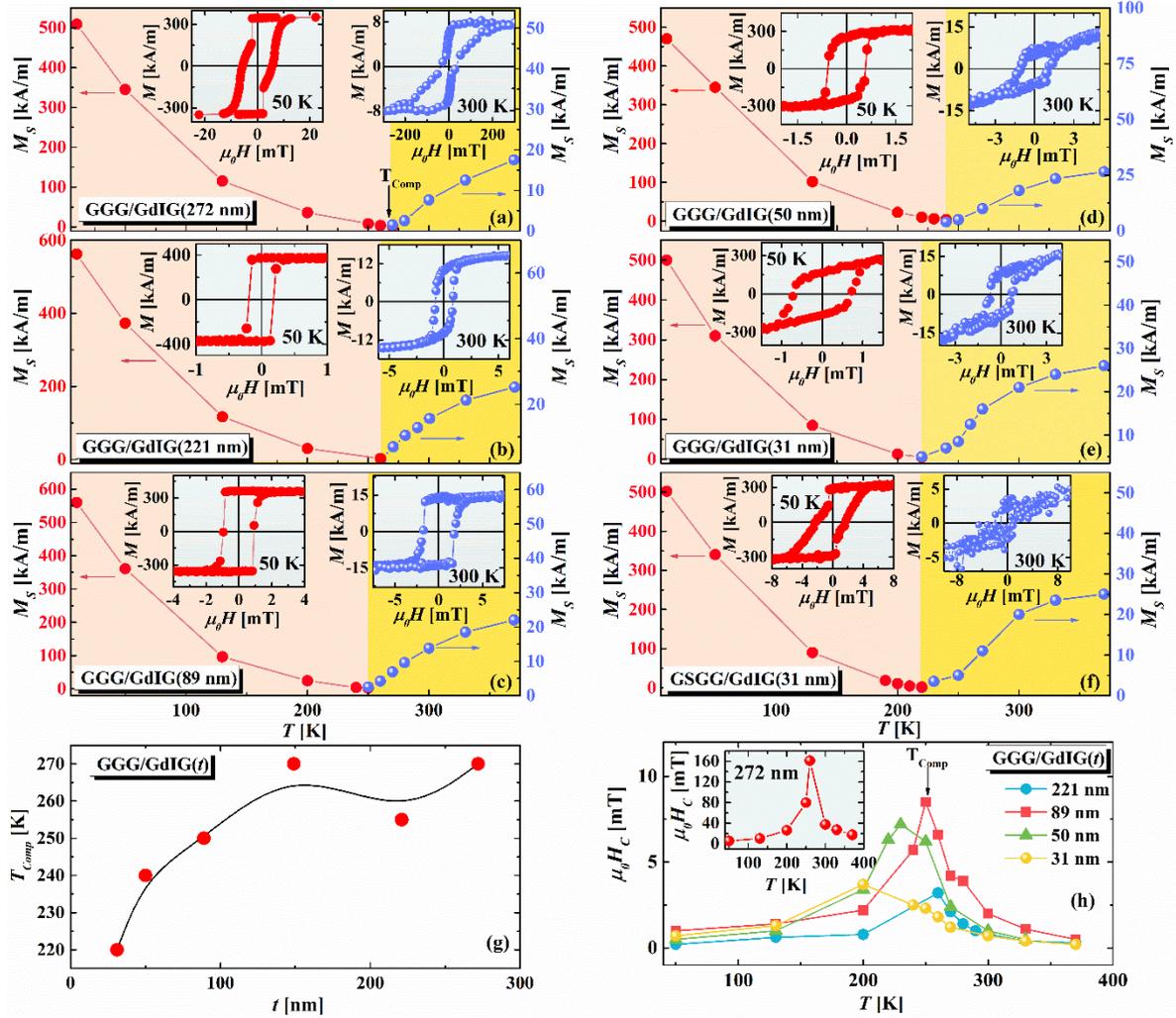

**Figure 2.** (a)-(f) Main panel: temperature dependence of saturation magnetization ($M_S$) for the G(S)GG/GdIG($t$) films with $t$ = 272-31 nm, left and right insets show $M(H)$ hysteresis loops for the corresponding heterostructures recorded at $T$ = 50 and 300 K, respectively. Note that the paramagnetic background of the substrate was subtracted from the $M(H)$ data after fitting a linear part at sufficiently high fields (above 300 mT) (g) Thickness dependence of the compensation temperature and, (h) temperature dependence of coercive field ($H_C$) for the GGG/GdIG films.

For all GdIG films, $M_S$ first decreases with decreasing temperature, becomes vanishingly small near the magnetic compensation temperature ($T_{Comp}$) and then increases drastically upon further decreasing temperature. While $T_{Comp}$ for the GGG/GdIG(*t*) films shifts down to low temperatures with decreasing thickness (*t*) of the film, as summarized in **Figure 2**(g), it remains invariant for the GdIG films grown on two different substrates (GGG and GSGG) with fixed thickness (*t* = 31 nm). The insets of **Figure 2**(a)-(f) show in-plane *M*(*H*) hysteresis loops measured at two selected temperatures, *T* = 300 and 50 K. All the GGG/GdIG(*t*) films exhibit strong in-plane (IP) anisotropy above and below $T_{Comp}$, whereas the GSGG/GdIG(31 nm) film shows out-of-plane (OOP) anisotropy around and above $T_{Comp}$, but IP anisotropy below 200 K. The OOP anisotropy for this film at *T* = 300 K was also confirmed by polar magneto-optical Kerr effect (p-MOKE) measurements (see **supplementary information: Figure S3**). The main panel of **Figure 2**(h) shows the coercive field, $H_C$ as a function of temperature obtained from the in-plane *M*(*H*) hysteresis loops for the GGG/GdIG(*t*) films for *t* = 221, 89, 50 and 31 nm. The temperature dependence of $H_C$ for the 272 nm film is shown separately in the inset of **Figure 2**(h). For all the films, $H_C$ increases drastically as $T_{Comp}$ is approached and exhibits a sharp peak at $T_{Comp}$, which is expected as $H_C$ for a compensated ferrimagnet like GdIG varies as $\frac{1}{|T - T_{Comp}|}$.[29] Another interesting fact is that at a fixed temperature, $H_C$ for the GSGG/GdIG(31 nm) film is more than twice of that for the GGG/GdIG(31 nm) film (see **supplementary information: Figure S5**), which is consistent with a recent report on the strain-induced enhancement of $H_C$ in REIGs[30].

Radio frequency (RF) transverse susceptibility (TS) measurements were employed to determine the temperature evolution of the effective magnetic anisotropy for both IP and OOP configurations. This TS technique can precisely detect the dynamic magnetic response of a magnetic material to a small and fixed amplitude RF magnetic field ($H_{RF}$) applied transverse to a static magnetic field ($H_{DC}$).[31] For a system with uniaxial anisotropy, the field dependence of TS shows sharp peaks at the anisotropy fields, $H_{DC} = \pm H_K$.[32] However, for a system with randomly dispersed magnetic anisotropy axes, TS exhibits cusps at the effective anisotropy fields, $H_{DC} = \pm H_K^{eff}$. We present the TS data as percentage change of transverse susceptibility defined as, $\frac{\Delta \chi_T}{\chi_T}(\%) = \frac{\chi_T(H_{DC}) - \chi_T(H_{DC}^{max})}{\chi_T(H_{DC}^{max})} \times 100\%$. Here, $\chi_T(H_{DC}^{max})$ is the value of transverse susceptibility at the maximum value of the applied dc magnetic field, $H_{DC}^{max}$ where, $H_{DC}^{max} \gg$ saturation field, $H_{DC}^{sat}$.

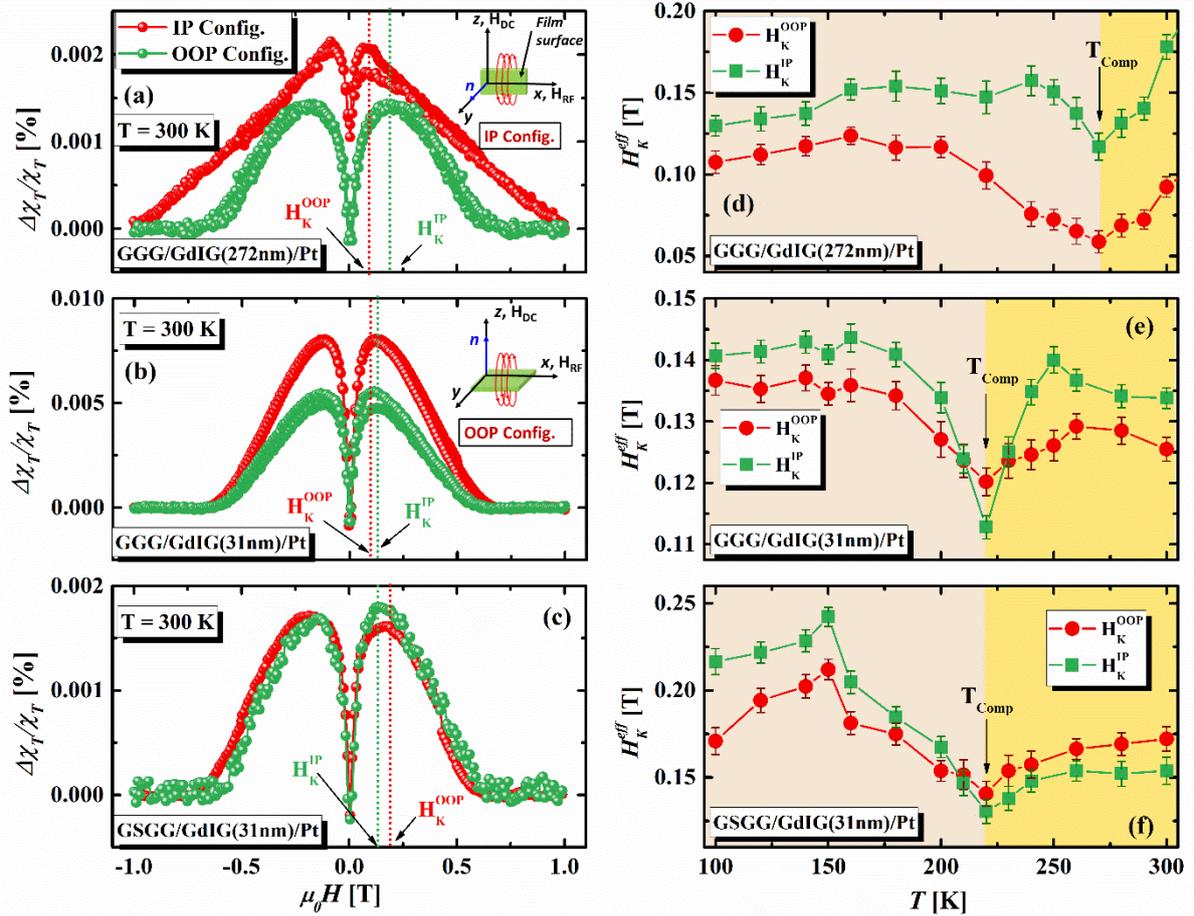

**Figure 3.** (a)-(c) IP (red) and OOP (green) transverse susceptibility data for bipolar field scans ( $+H_{DC}^{sat} \rightarrow -H_{DC}^{sat} \rightarrow +H_{DC}^{sat}$ ) for the G(S)GG/GdIG/Pt heterostructures with selected thicknesses of the GdIG films at $T$ = 300 K, (d)-(f) temperature dependence of the effective anisotropy fields ( $H_K^{eff}$ ) for both IP and OOP configurations for the corresponding heterostructures.

**Figure 3**(a)-(c) compares the bipolar field scans ($+H_{DC}^{max} \rightarrow -H_{DC}^{max} \rightarrow +H_{DC}^{max}$) of TS for our GdIG films with selected thicknesses for both IP ($H_{DC}$ lies along the film surface) and OOP ($H_{DC}$ is perpendicular to the film surface) configurations at $T$ = 300 K. For both IP and OOP configurations, $H_{DC} \perp H_{RF}$. For all these films, TS exhibits a maximum centering at the effective anisotropy fields: $\pm H_K^{eff}$ for both IP and OOP orientations. It is to be noted that the maxima observed in $\frac{\Delta \chi_T}{\chi_T}$ at $\pm H_K^{eff}$ are associated with the contributions from the spins pointing orthogonal to the direction of $H_{DC}$. In other words, for the OOP (IP) configuration, the TS scans probe the dynamics of the IP (OOP) spins.[33] Hence, we identify the effective anisotropy fields for the IP (OOP) configurations as $H_K^{eff} = H_K^{OOP}$ ($H_K^{IP}$). **Figure 3**(d)-(f) exhibits the temperature evolution of the IP and OOP anisotropy fields, *i.e.*, $H_K^{IP}(T)$ and $H_K^{OOP}(T)$, respectively in the temperature range 100 K ≤ $T$ ≤ 300 K. It is evident that $H_K^{IP} > H_K^{OOP}$ for all the GdIG(t)/GGG films above and below $T_{Comp}$, whereas, for the GdIG(31 nm)/GSGG film, $H_K^{IP} < H_K^{OOP}$ for $T \geq T_{Comp}$ but, $H_K^{IP} > H_K^{OOP}$ for $T < 200\,K$. This implies that higher (lower) field is required to rotate the IP (OOP) spins when the DC bias field is applied along the OOP (IP) directions for the GdIG(t)/GGG films.[33] In other words, the GdIG(t)/GGG films exhibit IP anisotropy for all temperatures. However, the magnetic easy axis for the GdIG (31 nm)/GSGG film transforms from OOP for $T \geq T_{Comp}$ to IP for $T < 200\,K$. This is consistent with our magnetometry results. Most importantly, both $H_K^{IP}(T)$ and, $H_K^{OOP}(T)$ exhibit a minimum around the compensation temperature. Decrease in the effective magnetic anisotropy constant close to the compensation temperature is common in rare-earth (RE) – transition metal (TM) based compensated ferrimagnets,[34][35][36][37] which is usually explained by canting of the sublattice magnetizations near $T_{Comp}$.

To understand the anomalous feature of $H_K^{IP}(T)$ and, $H_K^{OOP}(T)$ in the vicinity of $T_{Comp}$, let us first recall the magnetic structure of GdIG. The compensated ferrimagnetic insulator GdIG comprises of three magnetic sublattices; namely, the tetrahedrally coordinated $Fe^{3+}$ ions located at the *d*-sites, octahedrally coordinated $Fe^{3+}$ ions located at the *a*-sites and dodecahedrally coordinated $Gd^{3+}$ ions located at the *c*-sites.[22] Above $T_{Comp}$, the *c*-site $Gd^{3+}$ sublattice magnetization is small and the net magnetization is dominated by the *d*-site $Fe^{3+}$ sublattice, where the *d*-site and *a*-site $Fe^{3+}$ sublattices are strongly exchange coupled via antiferromagnetic (AF) interaction. With reducing temperature, the *c*-site $Gd^{3+}$ sublattice magnetization increases drastically. At $T_{Comp}$, the total magnetization of the *c*-site $Gd^{3+}$ sublattice and *a*-site $Fe^{3+}$ sublattice becomes equal in magnitude but antiparallel to the *d*-site $Fe^{3+}$ sublattice which makes the net magnetization zero. However, below $T_{Comp}$, the combined magnetization of the *c*-site $Gd^{3+}$ sublattice and *a*-site $Fe^{3+}$ sublattice overcomes the *d*-site $Fe^{3+}$ sublattice magnetization and thus, the net magnetization is dominated by the $Gd^{3+}$ sublattice. Since the magnetizations of the *c*-site $Gd^{3+}$ sublattice and *a*-site $Fe^{3+}$ sublattice are oriented along the same direction but both of them are anti-parallel to the *d*-site $Fe^{3+}$ sublattice magnetization, the three sublattices of GdIG can be reduced to two antiparallelly oriented effective sublattices for simplicity. Thus, the GdIG system can be treated as a two sublattice problem, namely, the *d*-site $Fe^{3+}$ sublattice (Fe-*d* sublattice) and the combined *c*-site $Gd^{3+}$ + *a*-site $Fe^{3+}$ sublattice (Gd-*c* + Fe-*a* sublattice). Following the approach of Sarkis et al.,[37] and Drzazga et al.,[35] the macroscopic effective magnetic anisotropy constant for our compensated ferrimagnetic GdIG films in the framework of a two-sublattice model can be expressed as,

$$K_{eff} = M_S^2 \left[ \frac{\lambda M_{Fe-d} \cdot M_{(Gd-c\,+\,Fe-a)} \{K_{eff}^{Fe-d} + K_{eff}^{(Gd-c\,+\,Fe-a)}\} + 2K_{eff}^{Fe-d} K_{eff}^{(Gd-c\,+\,Fe-a)}}{2\{K_{eff}^{Fe-d} M_{(Gd-c\,+\,Fe-a)}^2 + K_{eff}^{(Gd-c\,+\,Fe-a)} M_{Fe-d}^2\} + \lambda M_{Fe-d} \cdot M_{(Gd-c\,+\,Fe-a)} M_S^2} \right]. \quad (1)$$

Here, $\lambda$ is the strength of the inter-sublattice exchange interaction between the Fe-$d$ and Gd-$c$ + Fe-$a$ sublattices, $M_{Fe-d}$ and $M_{(Gd-c\,+\,Fe-a)}$ are the sublattice magnetizations and $K_{eff}^{Fe-d}$ and $K_{eff}^{(Gd-c\,+\,Fe-a)}$ are the macroscopic effective magnetic anisotropy constants for the Fe-$d$ and Gd-$c$ + Fe-$a$ sublattices, respectively, and $M_S = [M_{(Gd-c\,+\,Fe-a)} - M_{Fe-d}]$ is the net magnetization of the ferrimagnetic system with antiparallelly aligned sublattices. Equation 1 clearly indicates that $K_{eff} = 0$ at $T_{Comp}$, as $M_S = [M_{(Gd-c\,+\,Fe-a)} - M_{Fe-d}] = 0$, which qualitatively explains the observed minimum in both $H_K^{IP}(T)$ and, $H_K^{OOP}(T)$ at $T_{Comp}$.

## 2.3. Magnetic Field and Temperature Dependence of the Longitudinal Spin-Seebeck Effect

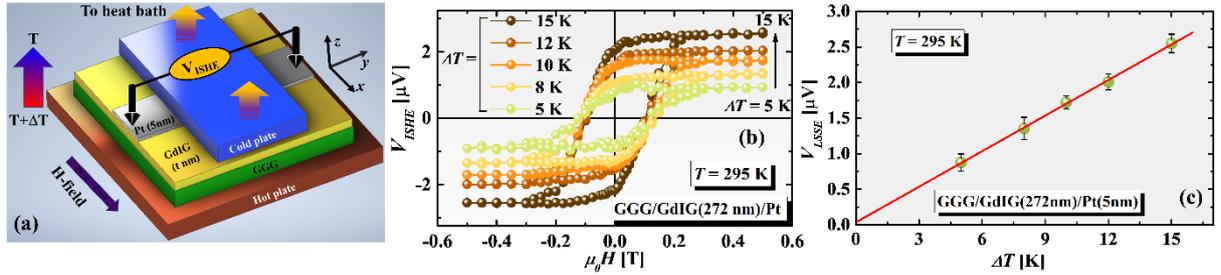

**Figure 4** (a) Schematic illustration of experimental configuration for the longitudinal spin Seebeck effect (LSSE) measurements on the G(S)GG/GdIG($t$)/Pt(5nm) heterostructures. (b) Magnetic field dependence of the ISHE-induced in-plane voltage, $V_{ISHE}(H)$ for different values of the temperature difference, $\Delta T$ for the GGG/GdIG (272 nm)/Pt heterostructure at $T$ = 295 K. (c) Background-corrected LSSE voltage, $V_{LSSE}(\mu_0 H_{sat}) = \frac{V_{ISHE}(+\mu_0 H_{sat}) - V_{ISHE}(-\mu_0 H_{sat})}{2}$ plotted against $\Delta T$, where $\mu_0 H_{sat}$ is the magnetic field required to saturate the GdIG magnetization.

Longitudinal spin Seebeck effect (LSSE) measurements on the GGG/GdIG/Pt and GSGG/GdIG/Pt heterostructures were performed using the longitudinal configuration where the sample is sandwiched between two copper plates, as illustrated in **Figure 4**(a). Each of these plates are equipped with a resistive chip-heater and Si-diode temperature sensor. We have used the same sample geometry for all the films including the distance between the contact leads ($L_y$ = 3 mm). A temperature gradient was applied along the +$z$-direction that generates a positive temperature difference, $\Delta T$ between the top (cold) and bottom (hot) plates. In this case the Pt layer is in contact with the cold plate whereas the G(S)GG substrate is in contact with the hot plate. The in-plane voltage generated along the *y*-direction in the Pt layer due to the ISHE ($V_{ISHE}$) was recorded as a function of the external in-plane magnetic field swept along the *x*-direction. Inside the magnetic insulator (MI) layer, GdIG, the thermally generated spin current is carried by the spin wave excitations or magnons.[38][39][40][41] In presence of the temperature gradient ($\vec{\nabla T}$) and an in-plane static magnetic field, a spatial gradient of magnon accumulation is developed inside the MI.[38][41][42] The magnon accumulation at the MI/Pt interface pumps the transverse spin current into the Pt layer.[41] The spin current density at the MI/Pt interface is given by, $\vec{J_S} = -\frac{b g_{eff}^{\uparrow\downarrow}}{a}\left[\frac{\cosh\left(\frac{t_{MI}}{l_{MI}}\right)-1}{\sinh\left(\frac{t_{MI}}{l_{MI}}\right)}\right] Q_{LSSE} \vec{\nabla T}$, where $g_{eff}^{\uparrow\downarrow}$ is the real part of the effective spin mixing conductance, $b = \frac{\gamma \hbar k_B T}{4\pi M_S} \frac{\int d^3k \left(\frac{\hbar\omega_k}{k_B T}\right) e^{\frac{\hbar\omega_k}{k_B T}} / \left(e^{\frac{\hbar\omega_k}{k_B T}}-1\right)^2}{\int d^3k\, e^{\frac{\hbar\omega_k}{k_B T}} / \left(e^{\frac{\hbar\omega_k}{k_B T}}-1\right)^2}$, $a = \frac{\hbar D_{MI}}{l_{MI}}$, $\hbar\omega_k$ is the magnon energy corresponding to the wave vector $\vec{k}$, $\gamma$ is the gyromagnetic ratio, $k_B$ is the Boltzmann constant, $D_{MI}$ and $l_{MI}$ are the diffusion coefficient and spin diffusion length of the MI, respectively and $Q_{LSSE}$ is the coefficient of LSSE.[41][43] Upon entering the Pt layer, the spin current transforms into a charge current via the ISHE. The density of the ISHE-induced charge current in the Pt layer is given by $\vec{J_C} = \left(\frac{2e}{\hbar}\right)\theta_{SH}^{Pt}(\vec{J_S} \times \vec{\sigma})$, where $\theta_{SH}^{Pt}$ is the spin Hall angle of Pt and $\vec{\sigma}$ is the spin-polarization vector.[44] Since the spin current, $\vec{J_S}$, at the interface between the magnetic and Pt layer diffuses into the Pt layer, integrating the charge current density, $\vec{J_C}$, along the *y* and *z* axes, one obtains the LSSE voltage across the Pt layer as [41][45][46]

$$V_{LSSE} = (R_y L_y \lambda_{Pt})\left(\frac{2e}{\hbar}\right)\theta_{SH}^{Pt}|J_S|\tanh\left(\frac{t_{Pt}}{2\lambda_{Pt}}\right). \quad (2)$$

Here, $R_y, L_y, \lambda_{Pt},$ and $t_{Pt}$ are the resistance between the contact leads on the Pt layer, the distance between the contact leads, the spin diffusion length, and the thickness of the Pt layer, respectively.[41] In **Figure 4**(b), we show the magnetic field dependence of the ISHE-induced in-plane voltage, $V_{ISHE}(H)$ for different values of the temperature difference between the hot ($T_{hot}$) and cold ($T_{cold}$) plates, $\Delta T = (T_{hot} - T_{cold})$ for the GGG/GdIG (272 nm)/Pt heterostructure at a fixed temperature $T = \frac{T_{hot}+T_{cold}}{2} = 295$ K. Clearly, $V_{ISHE}(H)$ traces well-defined hysteresis loop for all the $\Delta T$ values and the signal enhances upon increasing $\Delta T$. In **Figure 4**(c), we present the background-corrected longitudinal spin Seebeck effect voltage, $V_{LSSE}(\mu_0 H_{sat})$, defined as, $V_{LSSE}(\mu_0 H_{sat}) = \left[\frac{V_{ISHE}(\mu_0 H_{sat}) - V_{ISHE}(-\mu_0 H_{sat})}{2}\right]$ as a function of $\Delta T$, where $\mu_0 H_{sat}$ is the magnetic field required to saturate the GdIG magnetization. It is evident that the $V_{LSSE}$ signal scales linearly with $\Delta T$. Such linear $\Delta T$-dependence of the $V_{LSSE}$ signal actually reflects the intrinsic behavior of the LSSE signal originated from the thermally driven magnons in GdIG film, which is expected from **Equation 2**.

**Figure 5**(a)-(e) shows the $V_{ISHE}(H)$ hysteresis loops for the GGG/GdIG(*t*)/Pt(5nm) heterostructures with *t* = 272, 221, 89, 50, and 31 nm, respectively at few selected temperatures above and below their respective $T_{Comp}$ and, within the range 105 K $\leq T \leq$ 295 K for a fixed temperature difference between the hot and cold plates, $\Delta T$ = +10 K. The $V_{ISHE}(H)$ hysteresis loops for the GGG/GdIG(145 nm)/Pt heterostructure is shown in the **supplementary information (Figure S7)**. It is evident that the $V_{ISHE}(H)$ hysteresis loop changes its sign for all the five films (the six films including the 145 nm thickness) below their respective $T_{Comp}$. For example, as shown in the **Figure 5**(a), the $V_{ISHE}(H)$ loop for the 272 nm film ($T_{Comp}$ = 270 K) is regular for $T > 275 K$ but, becomes inverted for $T < 265 K$. Moreover, the $V_{ISHE}(H)$ signal strength decreases as $T_{Comp}$ is approached from high temperatures but, enhances upon

decreasing the temperature below $T_{Comp}$. **Figure 5**(f) exhibits $V_{ISHE}(H)$ hysteresis loops for the GSGG/GdIG(31nm)/Pt(5nm) heterostructure (having OOP anisotropy at room temperature which changes to IP anisotropy for $T < 200\ K$) at selected temperatures above and below $T_{Comp}$ = 220 K with a fixed temperature difference, $\Delta T$ = +10 K. Clearly, the $V_{ISHE}(H)$ hysteresis loop also changes its sign from regular for $T > 235\ K$ to inverted for $T < 215\ K$. Note that the $V_{ISHE}$ signal becomes too small to be detectable in the temperature range 220 K $\leq T < 235$ K.

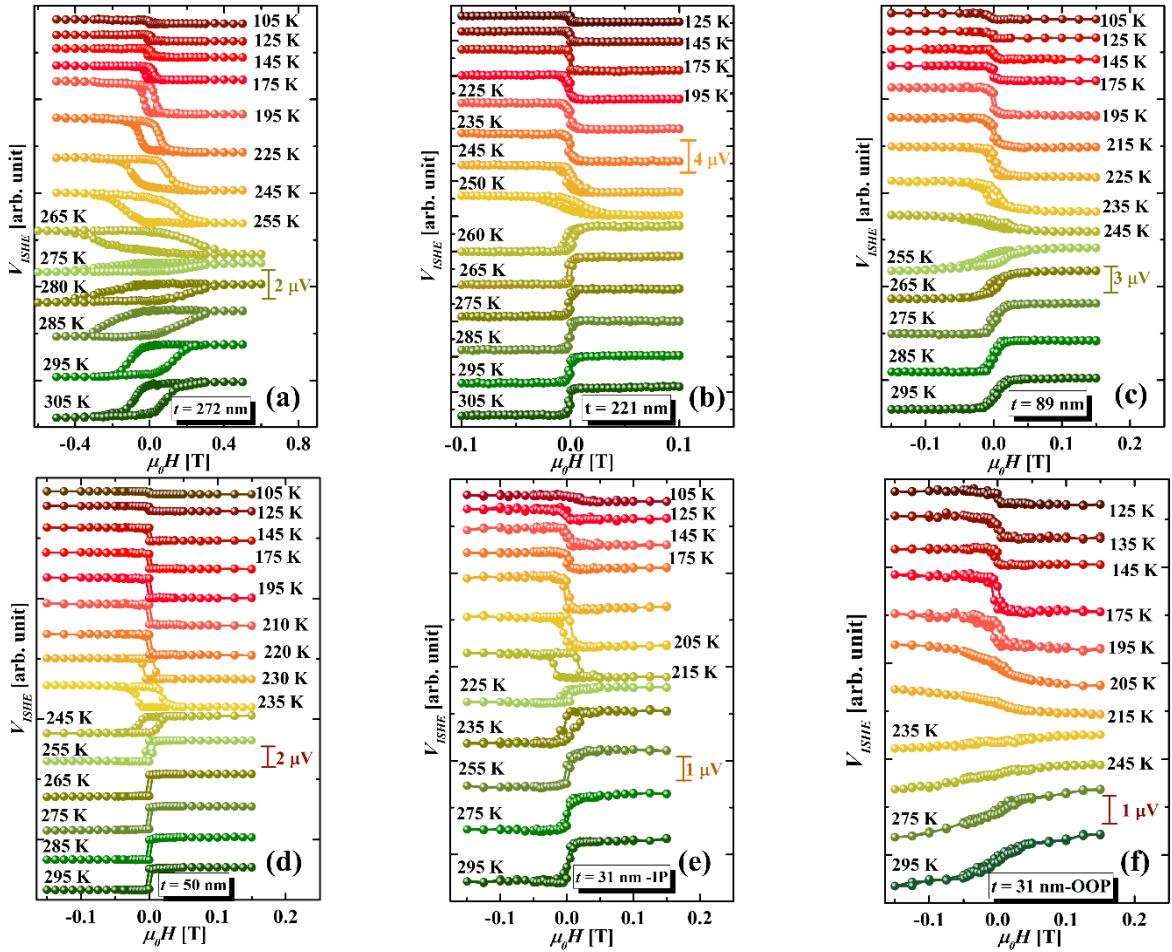

**Figure 5.** (a)-(e) $V_{ISHE}(H)$ hysteresis loops for the GGG/GdIG($t$)/Pt(5nm) heterostructures with $t$ = 272, 221, 89, 50 and 31 nm, respectively at few selected temperatures above and below their respective $T_{Comp}$ and, within the range 105 K $\leq T \leq$ 295 K for a fixed temperature difference between the hot and cold plates, $\Delta T$ = +10 K. (f) $V_{ISHE}(H)$ hysteresis loops for the GSGG/GdIG(31nm)/Pt(5nm) heterostructure at selected temperatures above and below $T_{Comp}$ = 220 K for $\Delta T$ = +10 K

To have a clearer insight, we have shown the two-dimensional *H-T* phase diagrams of $V_{ISHE}(H)$ isotherms in **Figure 6**(a)-(e) for the films with thicknesses *t* = 272, 221, 89, 50, and 31 nm, respectively, in the temperature range 105 K ≤ *T* < 295 K and for the $+\mu_0 H_{sat} \rightarrow -\mu_0 H_{sat}$ sweep, which undoubtedly reflect the sign reversal of the $V_{ISHE}(H)$ loops below their respective $T_{Comp}$. The asymmetric behavior of the *H-T* phase diagrams in the low field region signifies the hysteretic behavior of the $V_{ISHE}(H)$ loops for the GGG/GdIG(*t*)/Pt heterostructures.

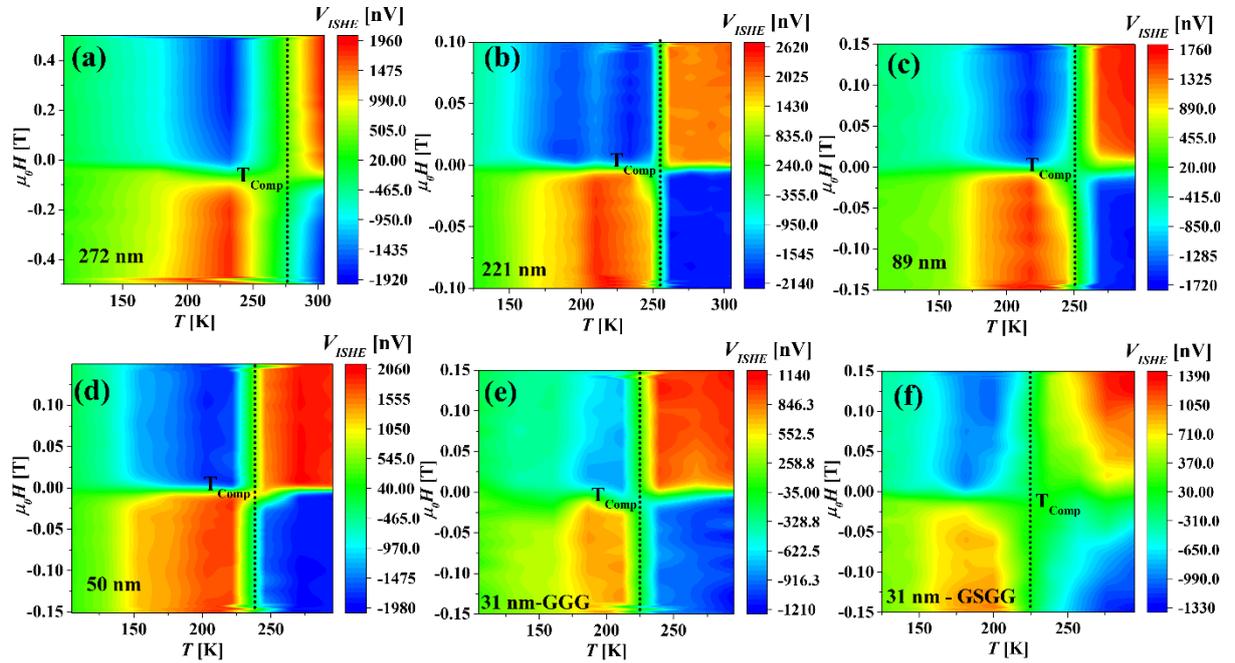

**Figure 6.** (a)-(f) two-dimensional *H-T* phase diagrams of $V_{ISHE}(H)$ isotherms for the G(S)GG/GdIG(*t*)/Pt(5nm) heterostructures with *t* = 272, 221, 89, 50, and 31 nm, respectively for the $+\mu_0 H_{sat} \rightarrow -\mu_0 H_{sat}$ sweep.

In **Figure 6**(f), we have shown the two-dimensional *H-T* phase diagram of $V_{ISHE}(H)$ isotherms for the GSGG/GdIG(31nm)/Pt(5nm) heterostructure within the aforementioned temperature range, which also reveals the sign reversal of the $V_{ISHE}(H)$ hysteresis loop below $T_{Comp}$ = 220 K. Comparing the *H-T* phase diagrams of $V_{ISHE}(H)$ for the GSGG/GdIG(31nm)/Pt heterostructure with those for the GGG/GdIG(t)/Pt heterostructures, it is evident that the sign reversal transition of the $V_{ISHE}$ signal across $T_{Comp}$ is sharper for the GGG/GdIG(t)/Pt heterostructures, *i.e.*, the sign change of the $V_{ISHE}$ signal occurs over a broader temperature range across $T_{Comp}$. This is possibly because of the fact that the $V_{ISHE}$ signal for the GSGG/GdIG(31nm)/Pt heterostructure becomes vanishingly small below 235 K, whereas for the GGG/GdIG(31nm)/Pt heterostructure, a detectable $V_{ISHE}$ signal was obtained even at 225 K. Moreover, the wide gap between the positive and negative maxima of the $V_{ISHE}(H)$ hysteresis loops in the *H-T* phase diagram of the GSGG/GdIG(31nm)/Pt heterostructure in the low field region clearly indicates the prolonged hysteretic behaviour of the $V_{ISHE}(H)$ loops.

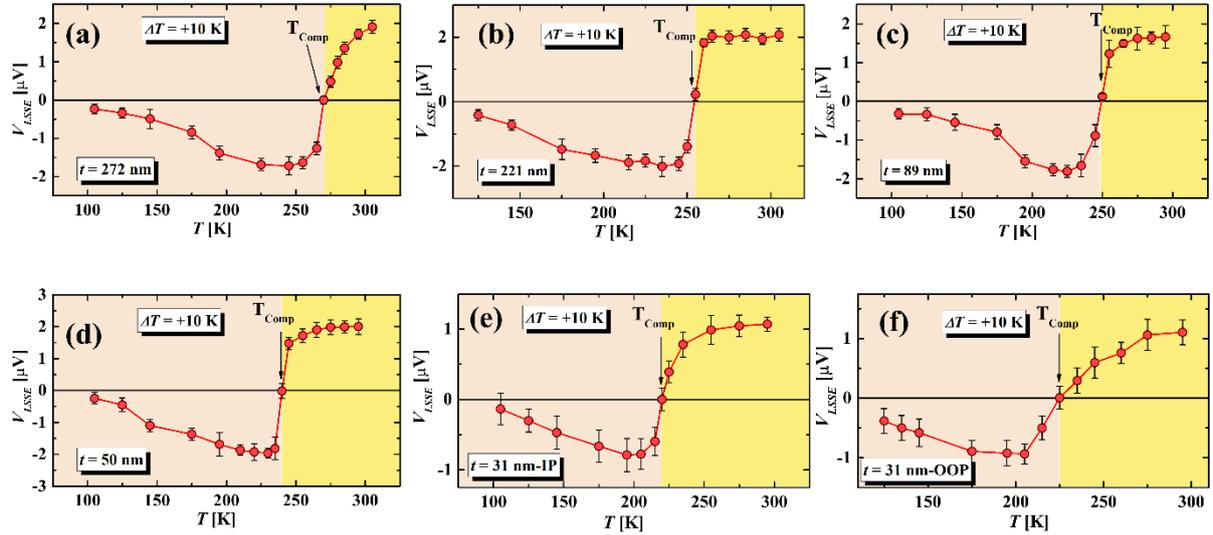

**Figure 7.** (a)-(f) Temperature dependence of the background-corrected longitudinal spin Seebeck effect voltage, $V_{LSSE}(T, \mu_0 H_{sat}) = \frac{V_{ISHE}(T, +\mu_0 H_{sat}) - V_{ISHE}(T, -\mu_0 H_{sat})}{2}$ for the G(S)GG/GdIG(t)/Pt heterostructures with *t* = 272, 221, 89, 50. and 31 nm, respectively in the temperature range 105 K ≤ *T* ≤ 295 K for *ΔT* = +10 K.

In **Figure 7**(a)-(f), we show the temperature dependence of the background-corrected longitudinal spin Seebeck effect voltage, $V_{LSSE}(T, \mu_0 H_{sat})$, defined as, $V_{LSSE}(T, \mu_0 H_{sat}) = \frac{V_{ISHE}(T, +\mu_0 H_{sat}) - V_{ISHE}(T, -\mu_0 H_{sat})}{2}$ for the G(S)GG/GdIG(t)/Pt heterostructures in the temperature range 105 K ≤ *T* ≤ 295 K for *ΔT* = +10 K, where $\mu_0 H_{sat}$ is the saturation magnetic field for our GdIG films. For the thickest film (*t* = 272 nm), $\mu_0 H_{sat}$ = 0.5 T whereas, for all other films, $\mu_0 H_{sat} \leq 0.15$ T. For all the films, the LSSE signal is positive for $T \geq T_{Comp}$ but becomes negative for $T < T_{Comp}$. Such "sign reversal transition" of the $V_{LSSE}$ signal in our compensated ferrimagnetic system in the vicinity of $T_{Comp}$ can be realized in terms of the reorientation of the sublattice magnetizations.[22][24] As we already discussed, the net magnetization of the GdIG film is dominated by the $d$-site $Fe^{3+}$ sublattice magnetization for $T \geq T_{Comp}$ whereas, the combined magnetization of the $c$-site $Gd^{3+}$ sublattice and $a$-site $Fe^{3+}$ sublattice overcomes the $d$-site $Fe^{3+}$ sublattice magnetization for $T < T_{Comp}$ and thus, the net magnetization is dominated by the $Gd^{3+}$ sublattice. Since the orientation of sublattice magnetizations govern the polarization of spin current, the reorientation of the sublattice magnetizations gives rise to the observed sign change in the $V_{LSSE}$ signal below $T_{Comp}$.[47] It is also important to note that the $V_{LSSE}$ signal approaches zero at low temperatures which is possibly due to the existence of a second sign reversal below 100 K arising from a competition between two magnon branches and exchange coupling at the GdIG/Pt interface.[22]

The fascinating sign reversal of the LSSE signal across $T_{Comp}$ in our GdIG films has enabled us to examine whether or not *a universal behavior exists for the temperature dependence of LSSE in compensated ferrimagnets.* We recall that a phenomenological universal scaling approach is often used to understand the critical phenomena and the nature of the second order magnetic phase transition of a ferromagnetic material via the magnetocaloric effect (MCE).[48][49][50] Here, we propose a similar scaling approach for the LSSE signal for our GdIG films near $T_{Comp}$. In our scaling approach, we normalized the $V_{LSSE}(T, \mu_0 H_{sat})$ signal with respect to its maximum value, $\frac{V_{LSSE}(T,\mu_0 H_{sat})}{|V_{LSSE}^{Max}|}$ on either side of $T_{Comp}$ for different thicknesses of our GdIG films and plotted this function against a rescaled temperature $\theta$. We define $\theta$ as,

$$\theta = \begin{cases} -(T - T_{Comp})/(T_{R1} - T_{Comp}), & \text{for } T \leq T_{Comp} \\ (T - T_{Comp})/(T_{R2} - T_{Comp}), & \text{for } T > T_{Comp}. \end{cases} \quad (3)$$

Here, $T_{R1}$ and $T_{R2}$ are the reference temperatures below and above $T_{Comp}$ which satisfy the condition, $\frac{|V_{LSSE}(T_{R1}, \mu_0 H_{sat})|}{|V_{LSSE}^{Max}(T \leq T_{Comp})|} = \frac{|V_{LSSE}(T_{R2}, \mu_0 H_{sat})|}{|V_{LSSE}^{Max}(T > T_{Comp})|} = $ constant ($n$), with $n = \frac{1}{2}$ for our present study. Note that the normalization factor(s), $|V_{LSSE}^{Max}(T \leq T_{Comp})|$ and $|V_{LSSE}^{Max}(T > T_{Comp})|$ are the maximum values of the $V_{LSSE}$ signal below and above $T_{Comp}$, respectively.

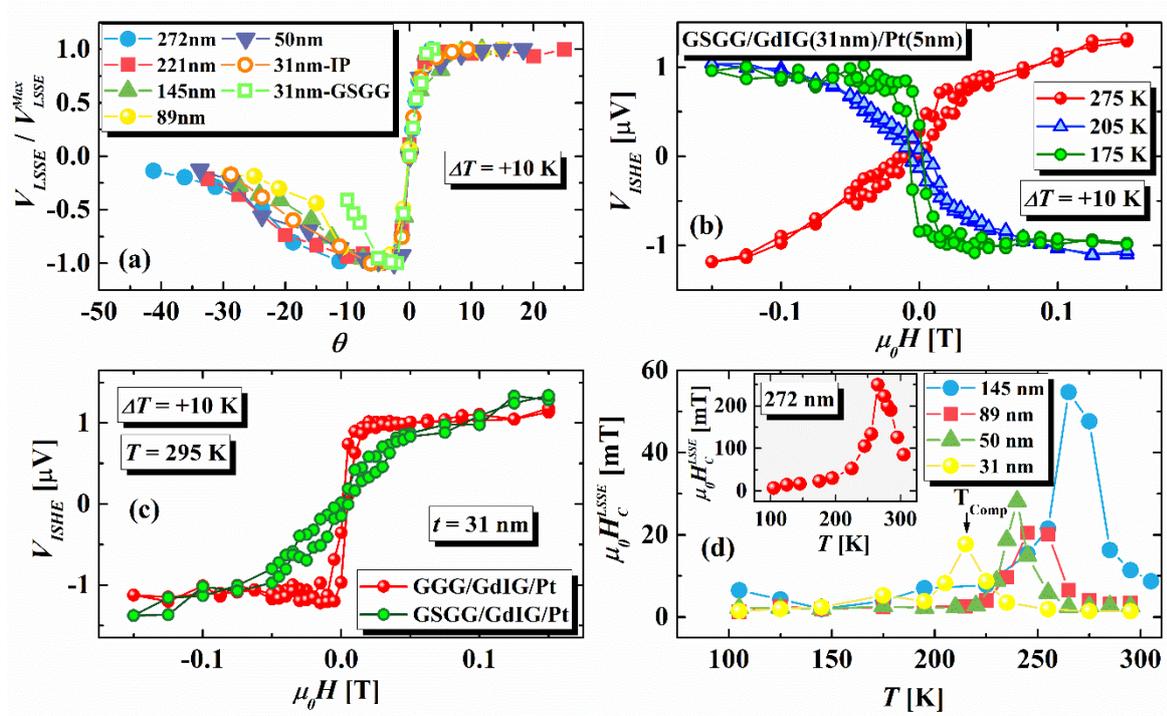

**Figure 8.** (a) $\frac{V_{LSSE}(T, \mu_0 H_{sat})}{|V_{LSSE}^{Max}|}$ vs. $\theta$ curves for the G(S)GG/GdIG(t)/Pt heterostructures with $t = $ 272, 221, 145, 89, 50, and 31 nm, respectively, (b) comparison of the $V_{ISHE}(H)$ hysteresis loops at $T = $ 275, 205 and 175 K with $\Delta T = +10$ K for the GSGG/GdIG(31nm)/Pt heterostructure, (c) comparison of the $V_{ISHE}(H)$ hysteresis loops at $T = $ 295 K with $\Delta T = +10$ K for the GSGG/GdIG(31nm)/Pt (OOP easy axis) and GGG/GdIG(31nm)/Pt (IP easy axis) heterostructures, (d) temperature dependence of the coercive field ($H_C^{LSSE}$) for the GGG/GdIG/Pt heterostructures.

As shown in **Figure 8**(a), the $\frac{V_{LSSE}(T,\mu_0 H_{sat})}{|V_{LSSE}^{Max}|}$ vs. $\theta$ curves for the GGG/GdIG(t)/Pt heterostructures with different thicknesses and hence, different $T_{Comp}$ nearly fall onto a single "master curve" in the vicinity of their respective $T_{Comp}$. This important finding highlights the universal behavior of the LSSE in compensated ferrimagnets. It should be noticed that the $\frac{V_{LSSE}(T,\mu_0 H_{sat})}{|V_{LSSE}^{Max}|}$ vs. $\theta$ curve for the GSGG/GdIG (31 nm)/Pt heterostructure slightly deviates from the master curve below $T_{Comp}$, mainly due to the different effect of magnetic anisotropy. It is noteworthy that the temperature-driven sign reversal transition for the GSGG/GdIG(31 nm)/Pt heterostructure (with OOP easy axis) is broader than that for the GGG/GdIG(31nm)/Pt heterostructures (with IP easy axis). In fact, the GGG/GdIG(50 nm)/Pt heterostructure shows the *sharpest* sign reversal transition in $V_{LSSE}(T,\mu_0 H_{sat})$, whereas the GSGG/GdIG (31 nm)/Pt heterostructure exhibits the *widest* sign reversal transition across $T_{Comp}$. We believe that the sharpness of the sign reversal transition of the $V_{LSSE}$ signal of our GdIG films across $T_{Comp}$ is strongly correlated to the magnetic anisotropy of our GdIG films. Our results indicate that GdIG is a promising candidate for thermally induced switching devices based on LSSE wherein, by tuning the device temperature across $T_{Comp}$, one can switch between positive and negative signals. The sharpness of this thermally induced switching is strongly correlated to the magnetic shape anisotropy of the GdIG film. Moreover, our 145 nm GGG/GdIG film has a rougher interface (RMS ~ 0.5 nm), a broader XRD peak and a lower saturation magnetization compared to other films (please see the **Supplementary information: Figures S1 and S4**), which is possibly due to the presence of a thin magnetically dead layer, or a not fully crystalline layer caused by slight thermal fluctuations during the film growth. These factors lead to the reduced LSSE signal in this film compared to other films. However, as seen from the **Figure 8**(a), the universal scaling behavior of the LSSE signal proposed in this work is not influenced by the crystallinity of the material. This means that our universal scaling model represents a general trend of the LSSE signal near the magnetic compensation for any compensated ferrimagnet.

In order to shed some light on the role of magnetic anisotropy on the LSSE signal, in **Figure 8**(b), we compare the $V_{ISHE}(H)$ hysteresis loops for the GSGG/GdIG(31nm)/Pt heterostructure recorded at three selected temperatures $T$ = 275, 205 and 175 K for $\Delta T$ = +10 K. At $T$ = 275 K, the system is far away from the magnetic compensation ($T_{Comp}$) and the $V_{ISHE}(H)$ hysteresis loop is elongated along the field axis; a typical feature observed in the $M(H)$ hysteresis loop recorded while scanning the applied magnetic field along the hard axis. Note that the GSGG/GdIG(31nm)/Pt heterostructure has an OOP easy axis at $T$ = 275 K and an IP magnetic field was swept during the LSSE measurement. Thus, the elongated $V_{ISHE}(H)$ hysteresis loop at $T$ = 275 K actually resembles the IP $M(H)$ loop. On the other hand, the $V_{ISHE}(H)$ hysteresis loop at $T$ = 205 K (just below $T_{Comp}$) is still elongated but shows comparatively sharper switching behaviour than at $T$ = 275 K. However, the $V_{ISHE}(H)$ hysteresis loop at $T$ = 175 K shows the sharpest switching behaviour and resembles the $M(H)$ hysteresis loop recorded while scanning the applied magnetic field along the magnetic easy axis. We may recall that the GSGG/GdIG(31nm)/Pt heterostructure shows IP easy axis below $T$ = 200 K which indicates that the $V_{ISHE}(H)$ hysteresis loop at $T$ = 175 K actually resembles the corresponding IP $M(H)$ loop. We also compare the $V_{ISHE}(H)$ hysteresis loops at $T$ = 295 K with $\Delta T$ = +10 K for the GSGG/GdIG(31nm)/Pt (OOP easy axis) and GGG/GdIG(31nm)/Pt (IP easy axis) heterostructures with same thickness of the GdIG film (31 nm) in **Figure 8**(c). Clearly, a distinct behaviour in the $V_{ISHE}(H)$ loops is evident, while the heterostructure with OOP magnetic easy axis shows a characteristic hard axis $V_{ISHE}(H)$ loop, the one with IP magnetic easy axis shows sharp switching behaviour in the $V_{ISHE}(H)$ loop and, bears a resemblance to the corresponding IP $M(H)$ hysteresis loop. These observations unambiguously highlight that the field dependence of the LSSE voltage is strongly susceptible to the orientation of magnetic anisotropy axis of our compensated ferrimagnetic GdIG films. Now let us correlate the temperature dependence of the LSSE signal (**Figure 7**) with the temperature dependence of effective anisotropy fields (**Figure 3**). While both $H_K^{IP}$ and $H_K^{OOP}$ increase just below $T_{Comp}$, the $V_{LSSE}$ signal changes its sign at the $T_{Comp}$ and increases in magnitude just below $T_{Comp}$ but, starts decreasing in magnitude gradually with further lowering the temperature. The decrease in the LSSE signal is possibly related to the increase in $H_K^{OOP}$ at low temperatures (below $T_{Comp}$).[19] Note that the magnetic anisotropy does not influence the magnetic configuration in the magnetically saturated state, *i.e.*, when all the spins are already aligned to the applied field direction. However, the propagation length ($\xi$) of the magnon spin current is influenced by the magnetic anisotropy. The magnon energy gap is related to the OOP effective anisotropy constant ($k_{eff}^{OOP}$) through the relation,[18][51] $\hbar\omega_M = (2k_{eff}^{OOP} + \mu_S B_z)$; where $\mu_S$ and $B_z$ are the permeability and applied magnetic field along the OOP direction (*z*-axis), respectively. The minimum value of the frequency dependent propagation length of the magnon spin current relates to the magnon energy gap through the expression,[52]

$$\xi_{Min} = \frac{a_0}{\sqrt{2}\alpha} \cdot \sqrt{\frac{J}{\hbar\omega_M}} = \frac{a_0}{\sqrt{2}\alpha} \cdot \sqrt{\frac{J}{(2k_{eff}^{OOP} + \mu_S B_z)}} \qquad (4)$$

where, $a_0$, $\alpha$, and $J$ are the lattice constant, the Gilbert damping parameter, and the strength of the Heisenberg exchange interaction between nearest neighbors, respectively. Since the magnetic field is applied along the IP direction during our LSSE measurements, $B_z = 0$. Therefore, **Equation 4** becomes $\xi_{Min} = \frac{a_0}{\sqrt{2}\alpha} \cdot \sqrt{\frac{J}{2k_{eff}^{OOP}}}$. Hence, higher OOP anisotropy increases the magnon energy gap, which leads to the propagation of only high frequency magnon spin current with shorter magnon diffusion length and thus, reduces the LSSE voltage. As mentioned earlier, the net spin current across the GdIG/Pt interface is determined by the spin injection efficiency of the $Gd^{3+}$ moment dominated gapless magnon mode ($\alpha$ mode) and the $Fe^{3+}$ moment dominated gapped magnon mode ($\beta$ mode).[22][24] As the $Gd^{3+}$ moment increases steeply upon lowering temperature compared to the $Fe^{3+}$ moment, the net spin current is dominated by the $\alpha$ mode magnons at low temperatures (below 80 K) but, it is dominated by the gapped $\beta$ mode magnons at higher temperatures. As $H_K^{OOP}$ increases rapidly below $T_{Comp}$, the magnon energy gap for the $\beta$ mode magnons also increases, which in turn reduces $V_{LSSE}$ for our GdIG/Pt films below $T_{Comp}$. However, decrease in $V_{LSSE}$ at lower temperatures (close to 100 K) is possibly due to the enhanced contribution of the $\alpha$ mode magnons which cancels out the contribution of the $\beta$ mode magnons.[22] The main panel of **Figure 8**(d) shows the coercive field, $H_C^{LSSE}$ as a function of temperature obtained from the $V_{ISHE}$ loops for the GdIG($t$)/GGG films with $t$ = 145, 89, 50, and 31 nm. The temperature dependence of $H_C^{LSSE}$ for the 272 nm film is shown separately in the inset of **Figure 8**(d). For all the films, $H_C^{LSSE}$ increases drastically as $T_{Comp}$ is approached and exhibits a sharp peak at $T_{Comp}$ and hence, mirrors the temperature dependence of $H_C$, as observed in **Figure 2**(h). This observation also indicates that $H_C^{LSSE}$ varies with $\frac{1}{|T-T_{Comp}|}$ for a compensated ferrimagnet.

Now let us discuss how the LSSE voltage varies as a function of thickness of the GdIG film at different temperatures. In presence of the temperature gradient, the thermally excited magnons propagate from the hotter region to the colder region along the direction of the temperature gradient. Such magnon propagation causes deviations of local magnetization from its equilibrium value which leads to an accumulation of magnons at the cold side of the temperature gradient.[52][14] The magnon spin current pumping from the magnetic insulator to the Pt layer is determined by the magnon accumulation, *i.e.*, the average number density of the magnons reaching the cold end of the temperature gradient. In other words, the LSSE voltage measured across Pt depends on the magnon accumulation at the cold end of the temperature gradient, which increases with increasing thickness of the magnetic insulator until a saturation value is reached.[14] The characteristic length scale for the saturation of the LSSE voltage signal is known as the mean magnon propagation length, $\langle \xi \rangle$. According to an atomistic spin model, the LSSE voltage is related to $\langle \xi \rangle$ through the expression,[14][17]

$$V_{LSSE}(t) = V_0 \left(1 - e^{-\frac{t}{\langle \xi \rangle}}\right) \quad (5)$$

Here, $V_0$ is the proportionality constant. **Equation 5** indicates that only thermally excited magnons will contribute to the magnon spin current and hence the LSSE signal which arrive at the cold end of the temperature gradient from a distance smaller than $\langle \xi \rangle$ from the GdIG/Pt interface. In other words, the saturation of the magnon spin current density at the GdIG/Pt interface and hence the saturation of LSSE voltage is achieved when the GdIG film thickness exceeds the mean propagation length of the thermally excited magnons. From the thickness dependence of $V_{LSSE}$ for our GGG/GdIG(*t*)/Pt heterostructures (please see **Supplementary information: Figure S8**), we have found that $V_{LSSE}$ at most of the temperatures increases with increasing thickness of the GdIG film, except for the temperature range of 275 – 225 K close to $T_{Comp}$ for the GdIG films with thickness ranging between *t* = 272 and 31 nm. However, it is evident from **Figure 8**(a) that the universal "thermal" scaling behavior of the LSSE voltage signal proposed in this work is not susceptible to the thickness dependence or the magnon propagation length scaling behavior of the LSSE voltage signal at least near the $T_{Comp}$.

In addition to the mean magnon propagation length, the frequency dependent magnon propagation length, $\xi(\omega)$, also influences the magnon accumulation at the cold side of the temperature gradient and hence the LSSE signal.[18] As we already discussed, the frequency dependent $\xi$ is strongly dependent on the magnetic anisotropy. However, for thin films, the film thickness also contributes to $\xi(\omega)$. Since the maximum value of $\xi(\omega)$, $\xi_{Max} \propto \frac{1}{(\hbar\omega)_{Min}}$,[18] high frequency magnons possess short $\xi$ and vice versa. Hence, for thicker films, both high frequency and low frequency magnons can reach the cold side of the thermal gradient and contribute to the LSSE signal. However, for thinner films, the low frequency magnons with large $\xi$ cannot contribute to the LSSE signal. For our GGG/GdIG(31nm)/Pt heterostructure, only high frequency magnons with shorter $\xi$ accumulate at the cold side of the temperature gradient and contribute to the LSSE signal, which explains the suppression of the LSSE signal for this film as compared to the thicker films.

Finally, let us elucidate the influence of the overall thermal resistance on the observed universal scaling behavior of the LSSE signal in our G(S)GG/GdIG/Pt heterostructures. In order to improve the thermal connectivity between the sample and hot/cold plates, cryogenic Apiezon N-grease was used. If a temperature difference $\Delta T$ is applied between the hot and cold plates, the effective temperature difference across the GdIG film, $\Delta T_{eff}$, becomes much smaller than $\Delta T$ because of the thermal resistance(s) offered by different regions during the measurement, namely, (1) the thermal grease layer between the cold plate and the Pt layer, (2) the Pt layer, (3) the GdIG/Pt interface, (4) the GdIG film, (5) the G(S)GG substrate and (6) the thermal grease layer between the G(S)GG substrate and the hot plate. Iguchi *et al*.[53] observed that the thermal resistance of a 10 μm-thick N-grease layer was substantially large and almost comparable to that of an yttrium iron garnet (YIG) slab at 300 K. They also found that the thermal resistance of the grease layer significantly influenced the temperature profile of LSSE signal especially at low temperatures when the thermal conductivity of the YIG was large. However, the thermal resistance of the Pt layer as well as the GdIG/Pt interface can be neglected.[53] So, the thermal grease layers and the G(S)GG substrates play dominating roles in consuming most of the applied temperature gradient. Estimation of the temperature profile of $\Delta T_{eff}$ as a function of temperature requires accurate knowledge of the temperature dependence of the thermal conductivity of GdIG. To maintain the same thermal contact conditions, we applied the same amount of N-grease between the sample and the hot/cold plates for all samples investigated. Therefore, the observed universal scaling behavior of the LSSE signal for our G(S)GG/GdIG/Pt heterostructures is free from the thermal resistance(s) of the N-grease layer and the G(S)GG substrate.

## 3. Conclusions

In summary, we have performed a comprehensive investigation of the longitudinal spin Seebeck effect in GGG/GdIG($t$)/Pt(5nm) heterostructures each of which possesses an in-plane magnetic easy axis, and the compensation temperature decreases from 270 to 220 K with decreasing film thickness from 272 to 31 nm, respectively. We found that the LSSE signal for all the heterostructures changes its sign below $T_{Comp}$. Using a proposed rescaling method, we have demonstrated a "universal scaling" behavior for the temperature dependence of LSSE signal for our GdIG films around their respective compensation temperatures. Furthermore, we have investigated LSSE in a 31 nm GdIG film grown on lattice-mismatched GSGG substrate that exhibits an out-of-plane magnetic easy axis at room temperature, but the magnetic easy axis changes its orientation to in-plane at low temperatures, which has been confirmed using radio frequency transverse susceptibility measurements. We have observed a clear distinction in the LSSE signal for the GSGG/GdIG(31 nm)/Pt heterostructure, relative to GGG/GdIG(31nm)/Pt showing an in-plane magnetic easy axis. Our findings underscore a strong correlation between the LSSE signal and the orientation of magnetic easy axis in compensated ferrimagnets.

## 4. Experimental Section/Methods

*Sample Preparation:* The gadolinium iron garnet (GdIG) thin films were grown by pulsed laser deposition (PLD), using a KrF excimer laser with a wavelength of 248 nm and a repetition rate of 2 Hz. The GdIG target was fabricated by mixing pure $Fe_2O_3$ and $Gd_2O_3$ powders. The mixed powders were pressed and sintered at 1200° C for 8 h in ambient atmosphere. Afterwards the target was cleaned inside the PLD chamber with more than $10^4$ pulses. The deposition conditions were calibrated to yield stoichiometric, single-crystalline thin films with smooth interfaces. The ideal film properties were achieved at elevated temperatures at 600°C, as measured by a thermocouple inside of the substrate holder, and at a deposition rate of 0.01 − 0.02 nm/s, while using an oxygen background atmosphere of 0.02 mbar. GdIG thin films with thicknesses between 31 nm and 272 nm were deposited on single-crystalline (111)-oriented GGG ($Gd_3Ga_5O_{12}$) and GSGG ($Gd_3Sc_2Ga_3O_{12}$) substrates of dimensions $5 \times 5 \times 0.5$ mm$^3$ and cooled down at a rate of approximately 5 K/min. The substrates were annealed for 8 h at 1200°C in oxygen atmosphere prior to the film deposition, to ensure a high substrate surface quality with a root-mean-square roughness below 0.2 nm. For the thin film garnets, the magnetic shape anisotropy governs the direction of the magnetic easy axis, favoring an alignment in the in-plane direction. Furthermore, lattice mismatch between the insulating iron garnet film and substrate induces in-plane compressive/tensile strain which leads to a magnetoelastic anisotropy favoring an out-of-plane (OOP) alignment of the magnetic easy axis.[54][55][56][57] The lattice constant of bulk GdIG is $a_{GdIG,bulk}$ = 1.2471 nm, whereas that for single crystalline GSGG substrate is $a_{GdIG,film}$ = 1.2554 nm.[58] Hence, the GdIG thin film grown epitaxially on a GSGG substrate exhibits an in-plane lattice constant >1.2471 nm, introducing tensile in-plane strain in the film, given by $\frac{a_{GdIG,bulk} - a_{GdIG,film}}{a_{GdIG,bulk}}$. Consequently, thin films grown on GGG strengthen the in-plane alignment of the magnetic easy axis, while for films grown on GSGG, magnetoelastic and magnetic shape anisotropy counteract each other, and perpendicular magnetic anisotropy (PMA) can be achieved at room temperature. After the GdIG thin film deposition, a $5 \times 2$ mm$^2$ Pt strip was prepared at room temperature by DC magnetron sputtering using a shadow mask. The films were annealed at 400°C for 30 minutes and subsequently cooled down to room temperature before Pt deposition to improve the GdIG/Pt interface quality.

*Structural Characterization:* The structural properties of the thin films were identified by X-ray diffraction (XRD) using monochromatic Cu K α radiation while the film surface morphology was investigated by atomic force microscopy (AFM).

*Magnetic Characterization:* The magnetic properties of the samples were analyzed by superconducting quantum interference device - vibrating sample magnetometry (SQUID-VSM) at temperatures between 10 and 370 K, and additionally by a polar magneto-optical Kerr effect (p-MOKE) setup operating at room temperature. The SQUID-VSM loops possess a magnetic contribution from both the film and the substrate. The paramagnetic signal of the substrate was removed by fitting a linear part at sufficiently high fields (above 300 mT) and then subtracting the linear background.

*Magnetic Anisotropy Measurement:* To understand the nature of the magnetic anisotropy in the G(S)GG/GdIG/Pt heterostructures, radio frequency (RF) transverse susceptibility (TS) measurements were carried out using a home-built self-resonant tunnel diode oscillator (TDO) circuit having a resonance frequency of ≈ 12 MHz with sensitivity of ~ 10 Hz. A physical property measurement system (PPMS) was utilized as a platform to sweep the external DC magnetic field ($H_{DC}$) and temperature. During the TS measurements, the films were firmly mounted inside an inductor coil (L) (which is a component of an LC tank circuit) and placed at the base of the PPMS sample chamber through a multi-purpose PPMS probe insert in such a way that the RF magnetic field ($H_{RF}$) generated inside L is parallel to the film surface, but perpendicular to the direction of $H_{DC}$. The amplitude of $H_{RF}$ is ~ 10 Oe. The remaining components of the TDO circuit were located outside the PPMS. In presence of $H_{DC}$, the dynamic susceptibility of the sample changes that in turns changes the inductance of L and gives rise to a shift in the resonance frequency of the LC tank circuit. TS as a function of $H_{DC}$ was obtained by recording the shift in the resonance frequency of the TDO oscillator circuit using an Agilent frequency counter while sweeping $H_{DC}$.

*Longitudinal Spin Seebeck Effect Measurement:* The longitudinal spin Seebeck effect (LSSE) on the G(S)GG/GdIG/Pt heterostructures was measured within the temperature range 105 K $\leq T \leq$ 295 K using a home-built set up assembled on a universal PPMS puck (see **Figure S6 in Supplementary information**). The heterostructures were sandwiched in between hot and cold plates both of which are made of copper. A thin layer of Kapton tape was thermally anchored to the bare surfaces of both the hot and cold plates. Cryogenic Apiezon N-grease was used to ensure good thermal connection between the sample surfaces and the Kapton tape layers anchored to the hot/cold plates. The Kapton tapes also facilitated to electrically insulate the cold (hot) plate from the top (bottom) surface of the heterostructures. An ultra-stable temperature difference ($\Delta T$) with $[\Delta T]_{Error} < \pm$ 2 mK between the hot and cold plates was achieved by individually controlling the temperatures of both these plates using two separate temperature controllers (Scientific Instruments Model no. 9700). The cold plate (top) was thermally connected to the base of the universal PPMS puck using Molybdenum screws. On the other hand, a 4 mm-thick Teflon block was sandwiched between the universal PPMS puck and the hot plate (bottom) to retain a temperature difference of ~ 10 K between them. In order to apply a temperature gradient along the +z-direction across the heterostructures, a Pt chip heater (PT-100 RTD sensors with 100 Ω resistance were used as heaters)[59] was attached to each of the hot and cold plates. To precisely control as well as sense the temperatures of both the plates, calibrated Si-diode sensors (DT-621-HR silicon diode sensors)[60] were attached to them. The sample temperature was recorded as, $T = T_{sample} = \left(\frac{T_{hot}+T_{cold}}{2}\right)$, where $T_{hot}$ and $T_{cold}$ correspond to the hot and cold plate temperatures, respectively. In presence of a stable temperature difference $\Delta T = (T_{hot} - T_{cold})$ across the heterostructure, thermally generated magnons will pump spin current from the GdIG layer to the adjacent Pt layer and the spin current is converted into charge current in the Pt layer via the ISHE. The ISHE voltage ($V_{ISHE}$) across the Pt layer of the G(S)GG/GdIG/Pt heterostructures was measured along the *y* direction using a Keithley 2182A nanovoltmeter, while sweeping a DC magnetic field produced by the superconducting magnet of the PPMS along the *x* direction. For the voltage measurements, two ultra-thin gold wires (25 µm diameter) were attached to the Pt strip using conducting silver paint (SPI supplies).

**Supporting Information**

Supporting Information is available from the Wiley Online Library or from the author.

**Acknowledgements**

Financial support by the US Department of Energy, Office of Basic Energy Sciences, Division of Materials Science and Engineering under Award No. DE-FG02-07ER46438 and by the German Research Foundation (DFG) within project AL618/37-1 is gratefully acknowledged.

**References**

[1] F. J. DiSalvo, *Science* **1999**, *285*, 703.

[2] K. Uchida, S. Takahashi, K. Harii, J. Ieda, W. Koshibae, K. Ando, S. Maekawa, E. Saitoh, *Nature* **2008**, *455*, 778.

[3] K. Uchida, H. Adachi, T. Kikkawa, A. Kirihara, M. Ishida, S. Yorozu, S. Maekawa, E. Saitoh, *Proc. IEEE* **2016**, *104*, 1946.

[4] D. Meier, D. Reinhardt, M. Van Straaten, C. Klewe, M. Althammer, M. Schreier, S. T. B. Goennenwein, A. Gupta, M. Schmid, C. H. Back, others, *Nat. Commun.* **2015**, *6*, 1.

[5] G. E. W. Bauer, E. Saitoh, B. J. Van Wees, *Nat. Mater.* **2012**, *11*, 391.

[6] Y. Kajiwara, K. Harii, S. Takahashi, J. Ohe, K. Uchida, M. Mizuguchi, H. Umezawa, H. Kawai, K. Ando, K. Takanashi, *Nature* **2010**, *464*, 262.

[7] J. Sinova, S. O. Valenzuela, J. Wunderlich, C. H. Back, T. Jungwirth, *Rev. Mod. Phys.* **2015**, *87*, 1213.

[8] K. Uchida, J. Xiao, H. Adachi, J. Ohe, S. Takahashi, J. Ieda, T. Ota, Y. Kajiwara, H. Umezawa, H. Kawai, *Nat. Mater.* **2010**, *9*, 894.

[9] J. Fontcuberta, H. B. Vasili, J. Gàzquez, F. Casanova, *Adv. Mater. Interfaces* **2019**, *6*, 1900475.

[10] J. Hirschner, M. Maryško, J. Hejtmánek, R. Uhreck\`y, M. Soroka, J. Burš\'\ik, A. Anadón, M. H. Aguirre, K. Kn\'\ižek, *Phys. Rev. B* **2017**, *96*, 64428.

[11] K. Uchida, T. Nonaka, T. Kikkawa, Y. Kajiwara, E. Saitoh, *Phys. Rev. B* **2013**, *87*, 104412.

[12] T. Niizeki, T. Kikkawa, K. Uchida, M. Oka, K. Z. Suzuki, H. Yanagihara, E. Kita, E. Saitoh, *AIP Adv.* **2015**, *5*, 53603.

[13] K. Uchida, H. Adachi, T. Ota, H. Nakayama, S. Maekawa, E. Saitoh, *Appl. Phys. Lett.* **2010**, *97*, 172505.

[14]   A. Kehlberger, U. Ritzmann, D. Hinzke, E.-J. Guo, J. Cramer, G. Jakob, M. C. Onbasli, D. H. Kim, C. A. Ross, M. B. Jungfleisch, *Phys. Rev. Lett.* **2015**, *115*, 96602.

[15]   K. Uchida, T. Kikkawa, A. Miura, J. Shiomi, E. Saitoh, *Phys. Rev. X* **2014**, *4*, 41023.

[16]   T. Kikkawa, K. Uchida, S. Daimon, Z. Qiu, Y. Shiomi, E. Saitoh, *Phys. Rev. B* **2015**, *92*, 64413.

[17]   E.-J. Guo, J. Cramer, A. Kehlberger, C. A. Ferguson, D. A. MacLaren, G. Jakob, M. Kläui, *Phys. Rev. X* **2016**, *6*, 31012.

[18]   U. Ritzmann, D. Hinzke, A. Kehlberger, E.-J. Guo, M. Kläui, U. Nowak, *Phys. Rev. B* **2015**, *92*, 174411.

[19]   V. Kalappattil, R. Das, M.-H. Phan, H. Srikanth, *Sci. Rep.* **2017**, *7*, 13316.

[20]   S. Lee, W. Lee, T. Kikkawa, C. T. Le, M. Kang, G. Kim, A. D. Nguyen, Y. S. Kim, N. Park, E. Saitoh, *Adv. Funct. Mater.* **2020**, *30*, 2003192.

[21]   V. Kalappattil, R. Geng, R. Das, M. Pham, H. Luong, T. Nguyen, A. Popescu, L. M. Woods, M. Kläui, H. Srikanth, *Mater. Horizons* **2020**, *7*, 1413.

[22]   S. Geprägs, A. Kehlberger, F. Della Coletta, Z. Qiu, E.-J. Guo, T. Schulz, C. Mix, S. Meyer, A. Kamra, M. Althammer, *Nat. Commun.* **2016**, *7*, 10452.

[23]   A. B. Harris, *Phys. Rev.* **1963**, *132*, 2398.

[24]   B. Yang, S. Y. Xia, H. Zhao, G. Liu, J. Du, K. Shen, Z. Qiu, D. Wu, *Phys. Rev. B* **2021**, *103*, 54411.

[25]   A. Yagmur, R. Iguchi, S. Geprägs, A. Erb, S. Daimon, E. Saitoh, R. Gross, K. Uchida, *J. Phys. D. Appl. Phys.* **2018**, *51*, 194002.

[26]   J. Cramer, E.-J. Guo, S. Geprägs, A. Kehlberger, Y. P. Ivanov, K. Ganzhorn, F. Della Coletta, M. Althammer, H. Huebl, R. Gross, others, *Nano Lett.* **2017**, *17*, 3334.

[27]   S. M. Zanjani, M. C. Onbasli, *AIP Adv.* **2019**, *9*, 35024.

[28]   D. Pesquera, X. Marti, V. Holy, R. Bachelet, G. Herranz, J. Fontcuberta, *Appl. Phys.*

*Lett.* **2011**, *99*, 221901.

[29] J. P. Hanton, A. H. Morrish, *J. Appl. Phys.* **1965**, *36*, 1007.

[30] H. Yamahara, B. Feng, M. Seki, M. Adachi, M. S. Sarker, T. Takeda, M. Kobayashi, R. Ishikawa, Y. Ikuhara, Y. Cho, others, *Commun. Mater.* **2021**, *2*, 1.

[31] H. Srikanth, J. Wiggins, H. Rees, *Rev. Sci. Instrum.* **1999**, *70*, 3097.

[32] A. Aharoni, E. H. Frei, S. Shtrikman, D. Treves, *Bull. Res. Counc. Isr.* **1957**, *6*, 215.

[33] R. P. Madhogaria, C.-M. Hung, B. Muchharla, A. T. Duong, R. Das, P. T. Huy, S. Cho, S. Witanachchi, H. Srikanth, M.-H. Phan, *Phys. Rev. B* **2021**, *103*, 184423.

[34] A. Ermolenko, *IEEE Trans. Magn.* **1976**, *12*, 992.

[35] Z. Drzazga, M. Drzazga, *J. Magn. Magn. Mater.* **1987**, *65*, 21.

[36] Z. Drzazga, *Phys. B+ C* **1985**, *130*, 305.

[37] A. Sarkis, E. Callen, *Phys. Rev. B* **1982**, *26*, 3870.

[38] S. S.-L. Zhang, S. Zhang, *Phys. Rev. B* **2012**, *86*, 214424.

[39] M. Johnson, R. H. Silsbee, *Phys. Rev. B* **1987**, *35*, 4959.

[40] C. Heide, *Phys. Rev. Lett.* **2001**, *87*, 197201.

[41] S. M. Rezende, R. L. Rodríguez-Suárez, R. O. Cunha, A. R. Rodrigues, F. L. A. Machado, G. A. F. Guerra, J. C. L. Ortiz, A. Azevedo, *Phys. Rev. B* **2014**, *89*, 14416.

[42] S. S.-L. Zhang, S. Zhang, *Phys. Rev. Lett.* **2012**, *109*, 96603.

[43] J. Xiao, G. E. W. Bauer, K. Uchida, E. Saitoh, S. Maekawa, *Phys. Rev. B* **2010**, *81*, 214418.

[44] E. Saitoh, M. Ueda, H. Miyajima, G. Tatara, *Appl. Phys. Lett.* **2006**, *88*, 182509.

[45] M. Arana, M. Gamino, E. F. Silva, V. Barthem, D. Givord, A. Azevedo, S. M. Rezende, *Phys. Rev. B* **2018**, *98*, 144431.

[46] A. Azevedo, L. H. Vilela-Leão, R. L. Rodr\'\iguez-Suárez, A. F. L. Santos, S. M. Rezende, *Phys. Rev. B* **2011**, *83*, 144402.

[47] Y. Ohnuma, H. Adachi, E. Saitoh, S. Maekawa, *Phys. Rev. B* **2013**, *87*, 14423.

[48] J. Y. Law, V. Franco, L. M. Moreno-Ramírez, A. Conde, D. Y. Karpenkov, I. Radulov, K. P. Skokov, O. Gutfleisch, *Nat. Commun.* **2018**, *9*, 2680.

[49] V. Franco, J. Y. Law, A. Conde, V. Brabander, D. Y. Karpenkov, I. Radulov, K. Skokov, O. Gutfleisch, *J. Phys. D. Appl. Phys.* **2017**, *50*, 414004.

[50] V. Singh, P. Bag, R. Rawat, R. Nath, *Sci. Rep.* **2020**, *10*, 6981.

[51] H. Jin, S. R. Boona, Z. Yang, R. C. Myers, J. P. Heremans, *Phys. Rev. B* **2015**, *92*, 54436.

[52] U. Ritzmann, D. Hinzke, U. Nowak, *Phys. Rev. B* **2014**, *89*, 24409.

[53] R. Iguchi, K. Uchida, S. Daimon, E. Saitoh, *Phys. Rev. B* **2017**, *95*, 174401.

[54] S. M. Zanjani, M. C. Onbaşlı, *J. Magn. Magn. Mater.* **2020**, *499*, 166108.

[55] A. Quindeau, C. O. Avci, W. Liu, C. Sun, M. Mann, A. S. Tang, M. C. Onbasli, D. Bono, P. M. Voyles, Y. Xu, *Adv. Electron. Mater.* **2017**, *3*, 1600376.

[56] O. Ciubotariu, A. Semisalova, K. Lenz, M. Albrecht, *Sci. Rep.* **2019**, *9*, 17474.

[57] J. J. Bauer, E. R. Rosenberg, S. Kundu, K. A. Mkhoyan, P. Quarterman, A. J. Grutter, B. J. Kirby, J. A. Borchers, C. A. Ross, *Adv. Electron. Mater.* **2020**, *6*, 1900820.

[58] H.-A. Zhou, L. Cai, T. Xu, Y. Zhao, W. Jiang, *Chinese Phys. B* **2021**.

[59] *Please see attached link for technical datasheet of PT-100 RTD sensors: https//docs.rs-online.com/2790/0900766b815bb28d.pdf*.

[60] *Please see attached link for technical datasheet of DT-621-HR silicon diode sensors: https//www.apo.nmsu.edu/arc35m/Instruments/ARCTIC/Development/ImagerPDR/LakeshoreTempCatalog_DT670.pdf*.

**Table of Contents (ToC)**

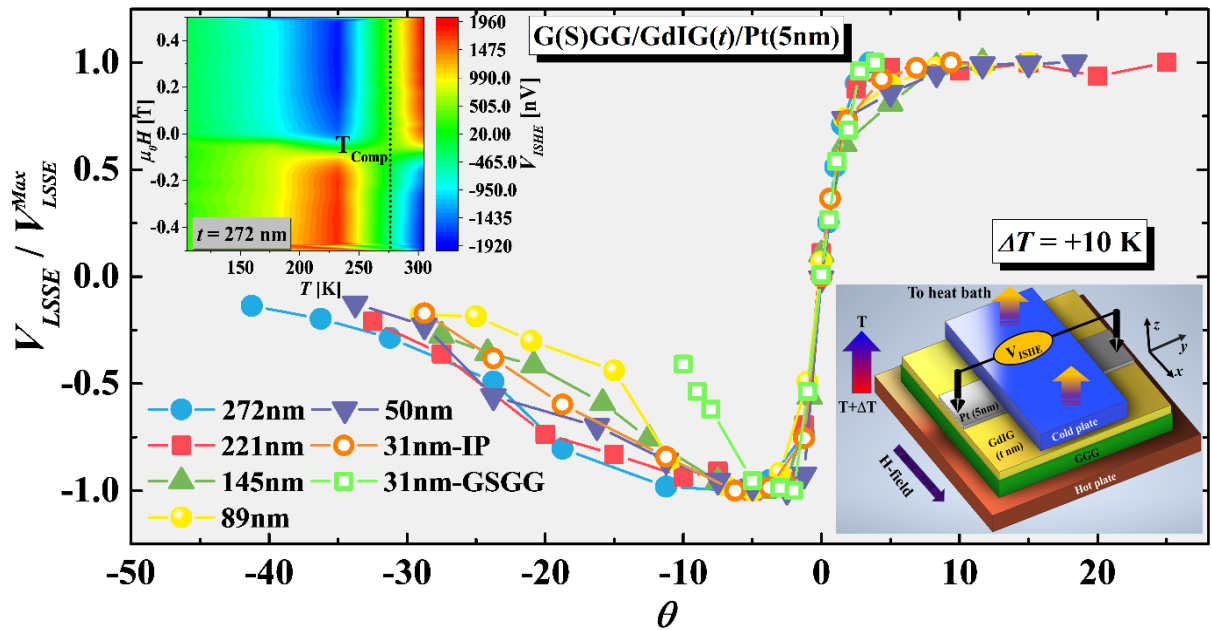

$Gd_3Fe_5O_{12}$ (GdIG) is a compensated ferrimagnetic insulator. We have demonstrated that the thermospin-voltage across the GGG/GdIG/Pt heterostructures with different GdIG thicknesses and hence, different compensation temperatures ($T_{Comp}$) fall onto a single master curve around their respective $T_{Comp}$. Such universal behavior would be helpful to fabricate novel spincaloritronic devices wherein the thermospin-voltage can be switched by appropriately tailoring the operational temperature.

# Scaling of the thermally induced sign inversion of longitudinal spin Seebeck effect in a compensated ferrimagnet: Role of magnetic anisotropy

*Amit Chanda, Noah Schulz, Manh-Huong Phan\* and Hari Srikanth\**

Department of Physics, University of South Florida, Tampa, Florida 33620, USA

E-mail: phanm@usf.edu; sharihar@usf.edu

*Christian Holzmann, Johannes Seyd and Manfred Albrecht\**

Institute of Physics, University of Augsburg, 86159 Augsburg, Germany

E-mail: manfred.albrecht@physik.uni-augsburg.de

**Supplementary Information**

**1. XRD and AFM images for the GdIG films with different thicknesses**

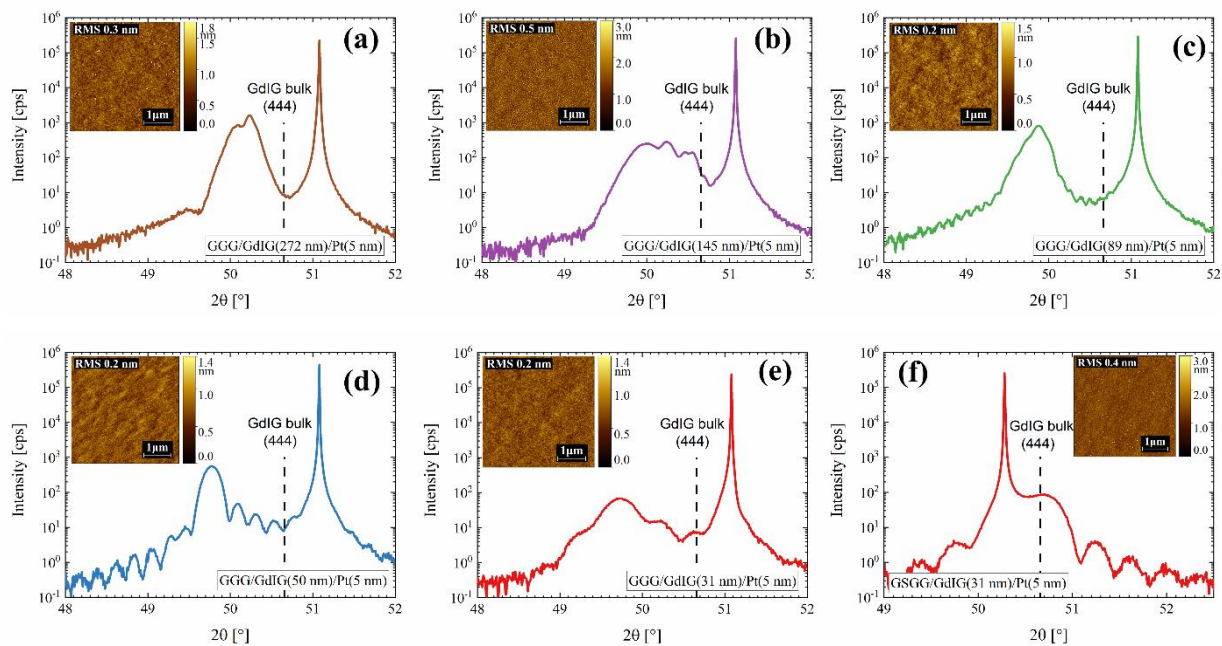

**Figure S1.** (a) – (f) Main panel: $\theta - 2\theta$ X-ray diffractograms (XRD) of the G(S)GG/GdIG ($t$)/Pt films with different film thickness $t$ ( = 272, 149, 89, 50 and 31 nm) and inset shows film morphology as visible in atomic force microscopy (AFM) for films grown on G(S)GG substrate.

## 2. XRD and AFM image for the 221 nm GdIG film

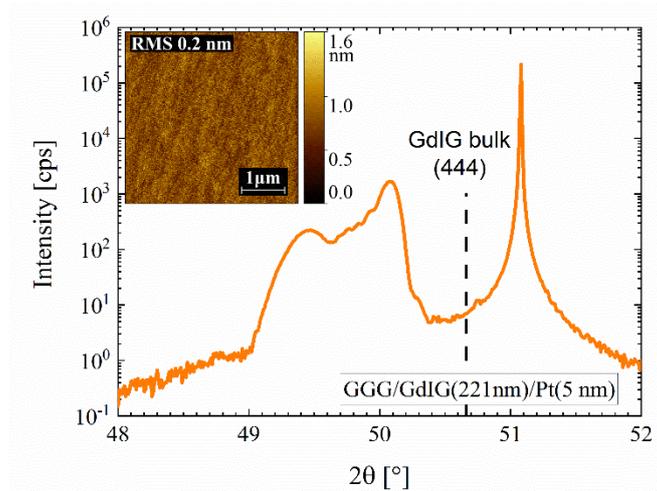

**Figure S2.** Main panel: $\theta - 2\theta$ X-ray diffractogram (XRD) of the GGG/GdIG(221nm)/Pt film and the inset shows film morphology as visible in an AFM image.

## 3. Comparison of IP M(H) and OOP Kerr loop for GSGG/GdIG (31nm)/Pt(5nm)

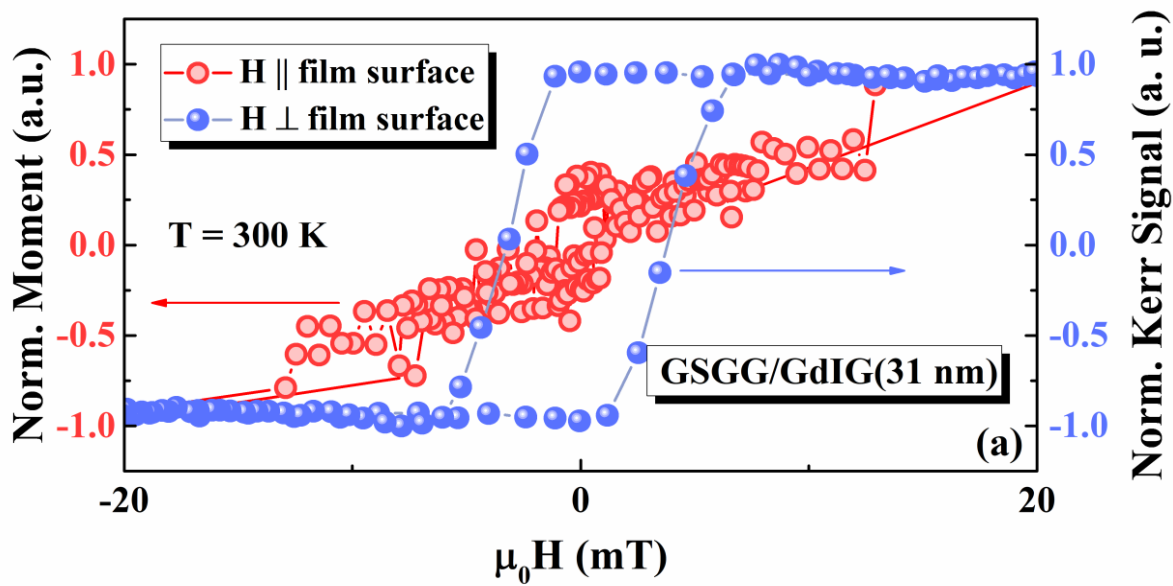

**Figure S3.** Left y-scale: normalized $M(H)$ loop for the GSGG/GdIG(31nm)/Pt heterostructure at $T$ = 300 K measured in the in-plane configuration and right y-scale: normalized polar magneto-optical Kerr effect (p-MOKE) signal as a finction of magnetic field applied along the OOP direction.

## 4. Temperature and magnetic field dependences of magnetization for GGG/GdIG(145nm)

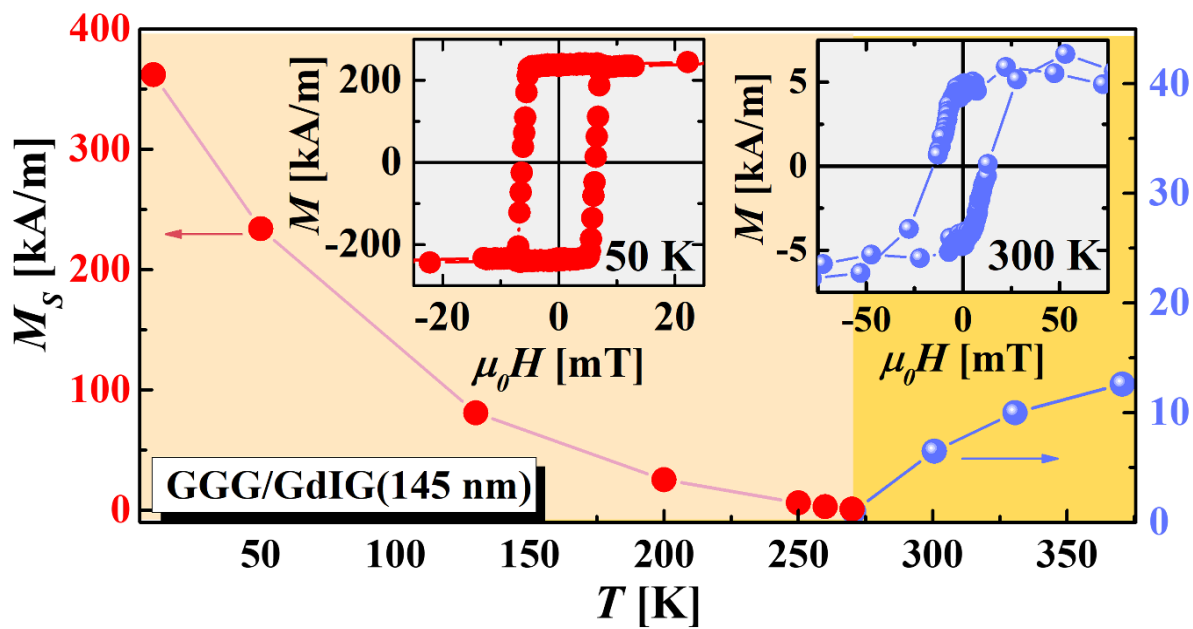

**Figure S4.** Main panel: temperature dependence of saturation magnetization ($M_S$) for the GGG/GdIG(145nm) heterostructure, left and right insets show $M(H)$ hysteresis loops for the corresponding heterostructure recorded at $T$ = 50 and 300 K, respectively.

## 5. Thickness dependence of M(H) loops for G(S)GG/GdIG(*t*) films

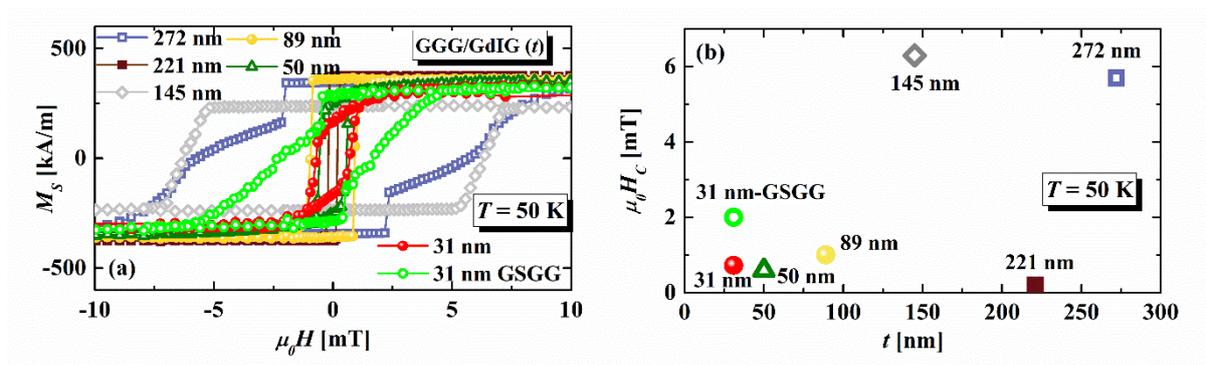

**Figure S5.** (a) Comparison of $M(H)$ hysteresis loops for G(S)GG/GdIG($t$) films at $T$ = 50 K and (b) thickness dependence of the coercive field for the corresponding heterostructures recorded at $T$ = 50 K.

## 6. Experimental configuration for LSSE measurements

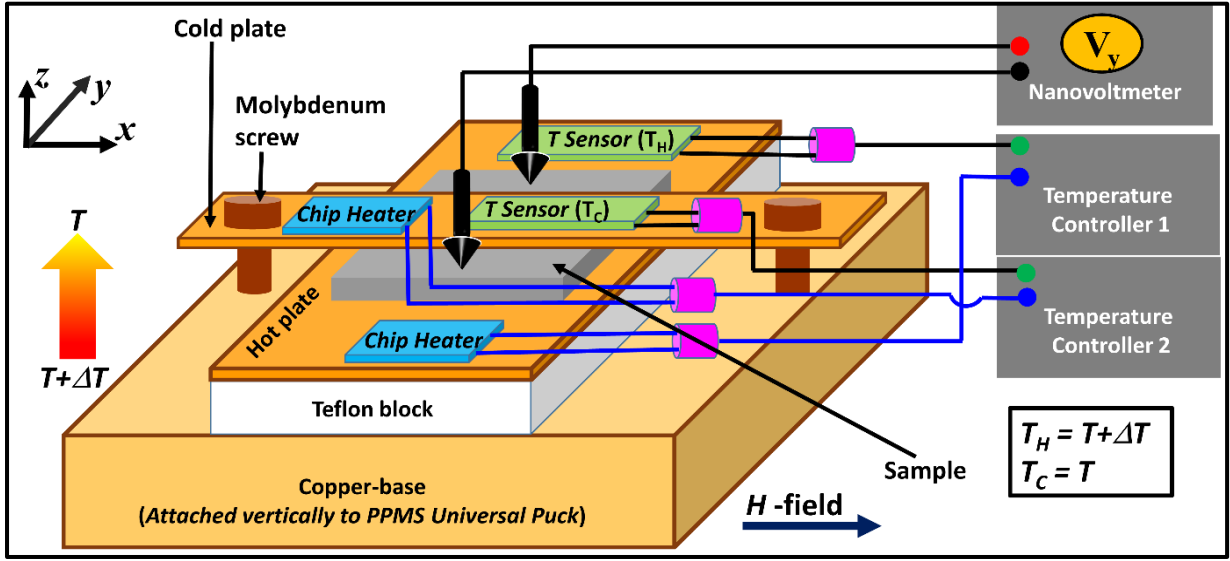

**Figure S6.** Schematic illustration of the experimental configuration for the longitudinal spin Seebeck effect (LSSE) measurements on the G(S)GG/GdIG($t$)/Pt(5nm) heterostructures.

## 7. Magnetic field and temperature dependence of LSSE signal for GG/GdIG(145nm)/Pt

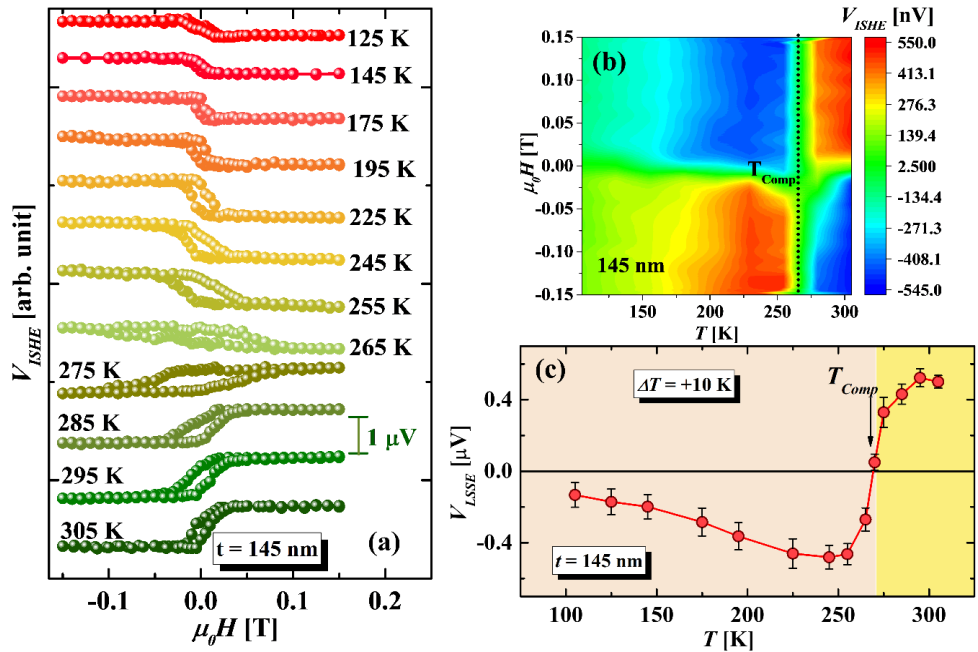

**Figure S7.** (a) $V_{ISHE}(H)$ hysteresis loop for the GGG/GdIG(145nm)/Pt(5nm) heterostructure at few selected temperatures above and below their respective $T_{Comp} = 270$ K and, within the range 105 K $\leq T \leq$ 295 K for a fixed temperature difference between the hot and cold plates, $\Delta T = +10$ K. (b) Corresponding two-dimensional $H$-$T$ phase diagrams of $V_{ISHE}(H)$ isotherms and (c) temperature dependence of the background-corrected LSSE voltage, $V_{LSSE}(T, \mu_0 H_{sat})$.

**8. Thickness dependence of the LSSE voltage at different temperatures for GGG/GdIG(*t*)/Pt(5nm)**

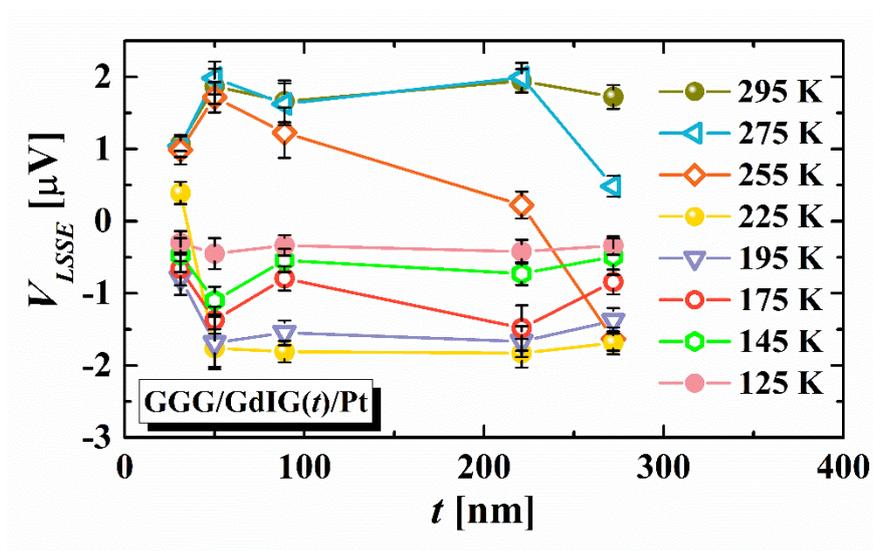

**Figure S8.** Thickness dependence of the LSSE voltage for the GGG/GdIG(*t*)/Pt(5nm) heterostructures at selected temperatures.